\documentclass[fleqn,usenatbib]{mnras}

\usepackage{newtxtext,newtxmath}
\usepackage{multicol}
\usepackage{xcolor}

\usepackage[T1]{fontenc}
\usepackage{blindtext}

\DeclareRobustCommand{\VAN}[3]{#2}
\let\VANthebibliography\thebibliography
\def\thebibliography{\DeclareRobustCommand{\VAN}[3]{##3}\VANthebibliography}

\usepackage{graphicx}	
\usepackage{amsmath}	
\usepackage{rotating}
\usepackage{tikz}

\title[Flare routine implementation into VULCAN]{The impact of time-dependent stellar activity on exoplanet atmospheres}

\author[A. J. Louca et al.]{
Amy J. Louca,$^{1}$\thanks{E-mail: louca@strw.leidenuniv.nl}
Yamila Miguel$^{2,1}$,
Shang-Min Tsai$^{3}$,
Cynthia S. Froning$^{4}$,
R. O. Parke Loyd$^5$,
\newauthor
Kevin France$^6$
\\
$^{1}$Leiden Observatory, Leiden University, Niels Bohrweg 2, 2333 CA Leiden, The Netherlands\\
$^{2}$SRON Netherlands Institute for Space Research, Niels Bohrweg 4, 2333 CA Leiden, the Netherlands\\
$^{3}$Atmospheric, Ocean, and Planetary Physics, Department of Physics, Oxford University, OX1 3PU, United Kingdom\\
$^{4}$McDonald Observatory, University of Texas at Austin, Austin, TX 78712 \\
$^5$School of Earth and Space Exploration, Arizona State University, Tempe, AZ 85287\\
$^6$Laboratory for Atmospheric and Space Physics, University of Colorado, 600 UCB, Boulder, CO 80309
}

\date{Accepted 2022 April 14. Received 2022 April 12; in original form 2021 December 24}

\pubyear{2021}

\begin{document}
\label{firstpage}
\pagerange{\pageref{firstpage}--\pageref{lastpage}}
\maketitle

\begin{abstract}

M-dwarfs are thought to be hostile environments for exoplanets. Stellar events are very common on such stars. These events might cause the atmospheres of exoplanets to change significantly over time. It is not only the major stellar flare events that contribute to this disequilibrium, but the smaller flares might also affect the atmospheres in an accumulating manner.
In this study we aim to investigate the effects of time-dependent stellar activity on the atmospheres of known exoplanets. We simulate the chemistry of GJ876c, GJ581c, and GJ832c that go from H$_2$-dominated to N$_2$-dominated atmospheres using observed stellar spectra from the MUSCLES-collaboration. We make use of the chemical kinetics code VULCAN and implement a flaring routine that stochastically generates synthetic flares based on observed flare statistics. Using the radiative transfer code petitRADTrans we also simulate the evolution of emission and transmission spectra. We investigate the effect of recurring flares for a total of 11 days covering 515 flares.  Results show a significant change in abundance for some relevant species such as H, OH, and CH$_4$, with factors going up to 3 orders of magnitude difference with respect to the preflare abundances. We find a maximum change of $\sim$12 ppm for CH$_4$ in transmission spectra on GJ876c. These changes in the spectra remain too small to observe. We also find that the change in abundance and spectra of the planets accumulate throughout time, causing permanent changes in the chemistry. We conclude this small but gradual change in chemistry arises due to the recurring flares.
\end{abstract}




\section{Introduction}

Stars are dynamic objects that evolve on both short- and long timescales. On the longer timescales, stars change their size and effective temperature, a process that can take billions of years and affect the atmospheres of orbiting planets (\citeauthor{Gallet2017} \citeyear{Gallet2017}; \citeauthor{Kislyakova2020} \citeyear{Kislyakova2020}).

On the other hand, stars also show energetic changes on shorter timescales due to the activity of their intense magnetic fields. Coronal mass ejections, stellar proton events, and stellar flares are the dynamical processes that can occur during such stellar activity. These dynamical processes are thought to have an impact on their planetary companions (e.g.\citeauthor{Segura2010} \citeyear{Segura2010}; \citeauthor{Venot2016} \citeyear{Venot2016}; \citeauthor{Tilley2018} \citeyear{Tilley2018}; \citeauthor{Chen2021} \citeyear{Chen2021}). 

M stars are particularly active stars (\citeauthor{Henry1994} \citeyear{Henry1994}; \citeauthor{Costa2006} \citeyear{Costa2006}), and extreme stellar activity will lead to sudden increases in received stellar flux and perhaps even severe compositional changes in the atmospheres  (e.g. \citeauthor{Segura2005} \citeyear{Segura2005}; \citeauthor{Zuluaga2014} \citeyear{Zuluaga2014}; \citeauthor{Rugheimer2015} \citeyear{Rugheimer2015}). The dissociation of molecules in to lighter particles due to intense stellar radiation might, in addition, reinforce atmospheric escape, and thus potentially losing considerable amounts of gas in the atmosphere. The question remains whether these compositional changes due to stellar activity and atmospheric loss are permanent or the initial equilibrium is restored.

\citet{Venot2016} focussed on this particular topic by using spectroscopic data of a flare event from AD Leo to simulate the atmospheres of two hypothetical hydrogen-dominated gas giants. They have found a post-flare steady state that is significantly different from the initial steady state, which could be detected in their spectra. In contrast, \citet{Segura2010} simulated water and ozone abundances of hypothetical exoplanet atmospheres in the Habitable Zone (HZ), where they also made use of flare data from AD Leo. They have found that there is no significant change in the abundances due to the UV-radiation emanating from the stellar flare. Similarly, \citet{Tilley2018} modelled the effects of repeated flaring on hypothetical Earth-like planets. They made use of time-resolved flare spectra from AD Leo and used flare occurence frequencies, with a varied proton event impact frequency. Their results show that the impact of repeated photon flare-events only have little effect on the ozone column depth. The multiple proton events, however, are able to rapidly destroy the ozone column. 
Building upon this research, \cite{Chen2021} recently showed that recurring flares from K- and M-dwarfs can significantly alter the chemical equilibria, using three-dimensional coupled chemistry-climate model simulations. \citet{Chen2021} focused on flares with energies greater than $10^{30}$ ergs and also included associated proton events. All these studies focused on the impact of stellar activity on exoplanet atmospheres, where atmospheric escape was not included in the calculations. 

In this study we aim to understand the longer-term effects of stellar activity and flares on atmospheric chemistry of rocky and gaseous planets. In contrast to previous studies, we include stellar flares that vary in size (i.e. small and big flares) by making use of synthetic light curves based on MUSCLES observations. Additionally, we include atmospheric escape in our model to see how this affects the compositions of the atmospheres. 
In particular, we study the evolution of both hydrogen- and nitrogen-dominated atmospheres for $\sim 11$ days. As case examples we adopt the planets GJ 876c, GJ 832c, and GJ 581c, of which the observed stellar spectra from their host stars are obtained from the MUSCLES collaboration (\citeauthor{France2016} \citeyear{France2016}, \citeyear{France2020}; \citeauthor{Youngblood2016} \citeyear{Youngblood2016}; \citeauthor{Loyd2016} \citeyear{Loyd2016}). 

The next section (\ref{sec:Methods}) explains the method in more detail, describing the numerical methods, stellar spectra, and the chosen planetary systems. Section \ref{sec:results} covers the resulting chemical compositions and emission and transmission spectra from the models. In section \ref{sec:discussion} we discuss the limitations of this study. Finally, we summarize and conclude the findings in section \ref{sec:conclusion}.

\section{Methods}
\label{sec:Methods}

\subsection{Planetary systems}

\begin{table*}
\centering
\begin{tabular}{c c c c c c c}
\hline
 & $M$ & $R$ & $T_{\mathrm{eff}}$  (K) & $a$ (AU) & $P$ (days) & Reference \\
 \hline
 \hline

GJ 876 & 0.33 $M_{\odot}$ & 0.36 $R_{\odot}$ & 3478 & - & - & \citeauthor{France2012} \citeyear{France2012} \\
GJ 832 & 0.45 $M_{\odot}$ & 0.48 $R_{\odot}$ & 3657 & - & - & \citeauthor{Bailey2009} \citeyear{Bailey2009}\\
GJ 581 & 0.31 $M_{\odot}$ & 0.299 $R_{\odot}$ & 3498 & - & - & \citeauthor{Bonfils2005} \citeyear{Bonfils2005}; \citeauthor{Bean2006} \citeyear{Bean2006}\\
\hline
GJ 876 c & 227 $M_{\oplus}$ & 14.01 $R_{\oplus}$ & 395 & 0.13 & 30.1 & \citeauthor{Marti2016} \citeyear{Marti2016}\\
GJ 832 c & 5.4 $M_{\oplus}$ & 1.68-2.03 $R_{\oplus}$ & 428 & 0.163 &  35.7 & \citeauthor{Wittenmyer2014} \citeyear{Wittenmyer2014}\\
GJ 581 c & 5.652 $M_{\oplus}$ & 1.7-2.08 $R_{\oplus}$ & 479 & 0.074 & 12.9 &\citeauthor{Trifonov2018} \citeyear{Trifonov2018}
\end{tabular}
\caption{Planet characteristics of the exoplanets used in this study. In contrast to the effective temperatures of the stars, which are obtained by fitting the black body emission function, the effective temperature of the planets in this study are calculated by multiplying the equilibrium temperature, assuming an albedo of zero, by a factor of $\sqrt{2}$ (see \citet{Moses2021}). 
}
\label{tab:planet_params}
\end{table*}

The planets studied in this paper are carefully selected considering a few criteria. First of all, we only consider planets that orbit stars from the MUSCLES survey (\citeauthor{France2016} \citeyear{France2016}; \citeyear{France2020}; \citeauthor{Youngblood2016}; \citeauthor{Loyd2016}). The stellar spectra from this survey give a more realistic representation of the UV-regime, as opposed to stellar spectral models based on blackbody emission that underestimates the received flux in the UV. Secondly, it is also important that the planets are not too close-in to the star, such that the temperatures are relatively low and we can assume that extreme hydrodynamic escape does not dominate (more on this in section \ref{tab:vertical_mixing}). We therefore only consider planets with orbital periods of $P$ > 10 days. A detailed description of the planets and their host stars can be found in table \ref{tab:planet_params}. Note that the radii of all three planets are not known and therefore we adopt the mass-radius relation for rocky 
\begin{equation}
\label{MR_rel}
    \frac{R}{R_{\oplus}} = (1.03\pm0.02)\left(\frac{M}{M_{\oplus}}\right)^{0.29\pm0.01}
\end{equation}
 and gaseous exoplanets
 \begin{equation}
 \label{MR_rel_gas}
     \frac{R}{R_{\oplus}} = (0.70 \pm 0.11)\left(\frac{M}{M_{\oplus}}\right)^{0.63\pm0.04}
 \end{equation}
  for $M_p \leq 120 \mathrm{\ } M_{\oplus}$ (\citeauthor{Otegi2019} \citeyear{Otegi2019}). For $M_p > 120 \mathrm{\ } M_{\oplus}$ we use the relation,
  \begin{equation}
      \frac{R}{R_{\oplus}} = 17.78\left(\frac{M}{M_{\oplus}}\right)^{-0.044}
  \end{equation}  
  as stated in \citet{Chen2017}. 

\subsubsection*{GJ 581}

GJ 581 is a spectral type M3V star at a distance of 6.21 pc from our solar system. It is known to host at least 3 planets (\citeauthor{Udry2007} \citeyear{Udry2007}), however some observational data favours 4-6 exoplanets (\citeauthor{Mayor2009} \citeyear{Mayor2009}; \citeauthor{Vogt2019} \citeyear{Vogt2019}, \citeyear{Vogt2012}; \citeauthor{Escude2010} \citeyear{Escude2010}; \citeauthor{Tuomi2011} \citeyear{Tuomi2011}, \citeyear{Tuomi2012}; \citeauthor{Hatzes2013} \citeyear{Hatzes2013}; \citeauthor{Robertson2014} \citeyear{Robertson2014}). 
Relatively close to the host star we have GJ 581 c - a warm planet outside the habitable zone. We assume that the atmosphere of GJ 581 c is dynamically stable and no hydrodynamical escape occurs. GJ 581 c then becomes a perfect candidate to investigate further.
With a mass of 5.4$M_{\oplus}$ and an estimated radius of 1.7-2.08$R_{\oplus}$ (from eq. \ref{MR_rel}-\ref{MR_rel_gas}), GJ 581c is at the border of being a Neptune-like planet or an Earth-like planet. It is therefore not evident whether we deal with a hydrogen or nitrogen-dominated atmosphere and thus both cases are investigated in this study. 

\subsubsection*{GJ 876}

At a distance of merely 4.7 pc this spectral type M4V is known to host 4 gas giant-like planets (\citeauthor{Delfosse1998} \citeyear{Delfosse1998}; \citeauthor{Marcy1998} \citeyear{Marcy1998}; \citeyear{Marcy2001}; \citeauthor{Rivera2005} \citeyear{Rivera2005}; \citeauthor{Rivera2010} \citeyear{Rivera2010}).
The planet companions are thought to consist of two Neptune-like planets (d and e) and two Jupiter-like planets (b and c). Two planets are of particular interest to this study due to their position in the habitable-zone, i.e. GJ 876 b ($\sim$ 0.21 AU) and GJ 876 c ($\sim$ 0.13 AU) (\citeauthor{Trifonov2018} \citeyear{Trifonov2018}). Due to the relatively large size of GJ 876 b, it is not expected that it is much affected by thermal atmospheric escape, also considering its distance to the host star. GJ 876 c, on the other hand, is somewhat smaller in size (although still quite large, with R $\approx$ 1.25 $R_J$ and M $\approx$ 0.714 $M_{J}$) and closer to the star, which may result in stronger thermal atmospheric escape and is therefore the chosen planet for this study.

\subsubsection*{GJ 832}

Another relatively close-by star hosting planets is the spectral type M1.5 star GJ 832, with a distance of 4.93 pc. This star is known to host a Jupiter-like planet and a super-Earth (\citeauthor{Bailey2009} \citeyear{Bailey2009}; \citeauthor{Wittenmyer2014} \citeyear{Wittenmyer2014}). The super-Earth, GJ 832 c, has an orbital period of $\sim$ 35.7 days, placing it within the habitable zone. Its minimum mass of $m\sin(i) \approx 5.4 \mathrm{\ M_{\oplus}}$ lies just at the border between super-Earth like planets or Neptune-like planets (\citeauthor{Otegi2019} \citeyear{Otegi2019}). We therefore evaluate both scenarios. The radius of this planet is still unknown and we therefore use eqs. \ref{MR_rel} - \ref{MR_rel_gas}, giving a radius between 1.68 - 2.03 $R_{\oplus}$. 

\subsection{Stellar Spectra}

\subsubsection{Quiescent Spectral Energy Distributions}

\begin{figure*}
    \centering
    \includegraphics[scale=0.6]{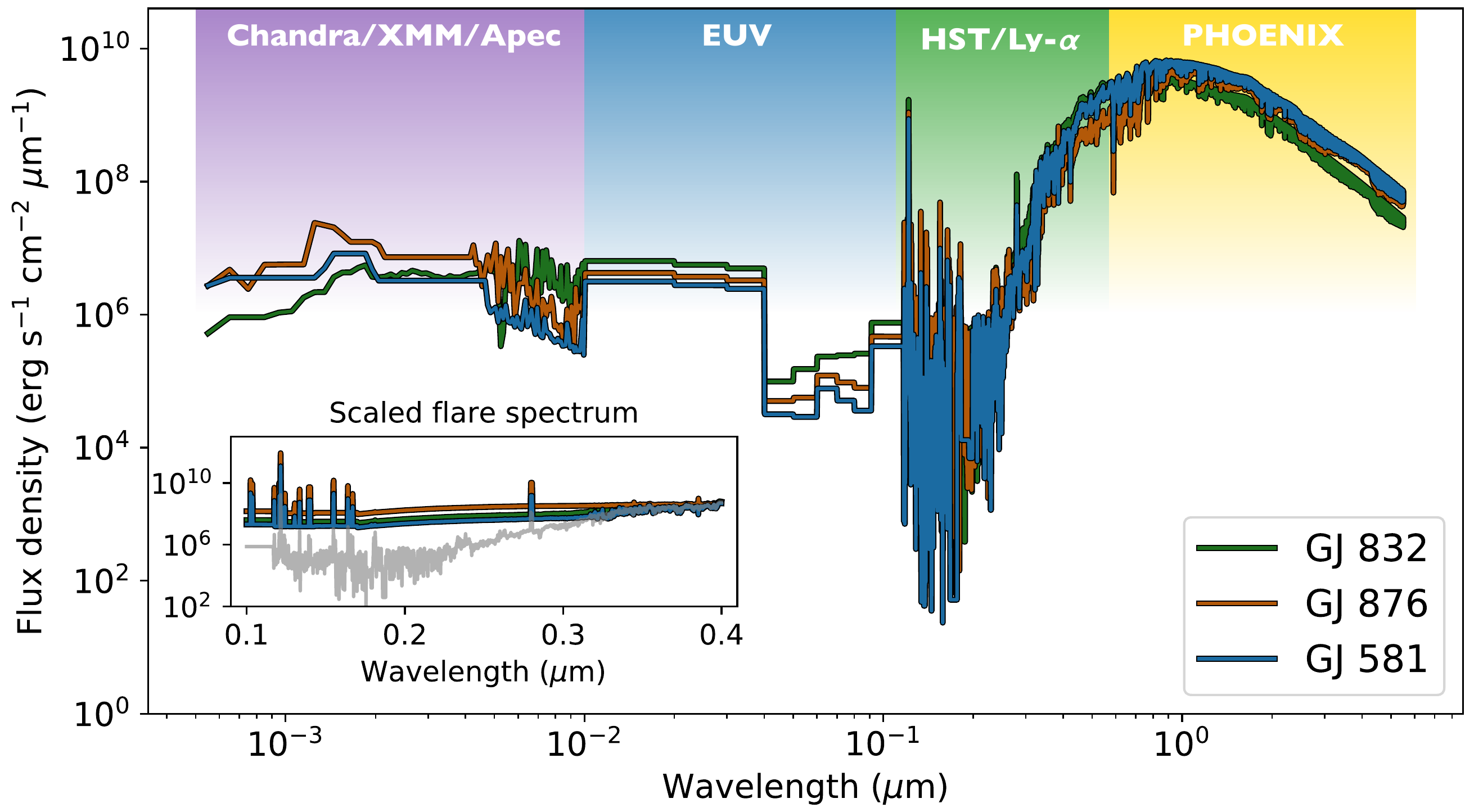}
     \caption{Quiescent spectra (main figure) of GJ 832, GJ 876 and GJ 581 from the MUSCLES survey, as scaled to their surface flux. The flare spectra in the UV-regime (0.1 - 0.4 $\mu$m) are shown in the box-figure for GJ 832 (green), GJ 876 (orange), and GJ 581 (blue), which all have an intensity of 100x quiescent values and are scaled accordingly to the quiescent Si IV doublet (by integrating synthetic and quiescent spectra over 0.139-0.1405$\mu$m). The grey spectrum in the box-figure represents the quiescent spectrum of GJ 832 from the main figure as an example to show the difference between the quiescent- and flare spectra. The colored regions depict the different models/observations used for that wavelength range.}
    \label{fig:spectra}
\end{figure*}

Quiescent stellar spectra are obtained from the Measurements of the Ultraviolet Spectra Characteristics of Low-mass Exoplanetary Systems (MUSCLES) collaboration (\citeauthor{France2016} \citeyear{France2016}, \citeyear{France2020}; \citeauthor{Youngblood2016} \citeyear{Youngblood2016}; \citeauthor{Loyd2016} \citeyear{Loyd2016}). The MUSCLES survey is a selection of observations of M- and K-dwarf exoplanet host stars in the optical, UV, and X-ray regime. The data is obtained using the Hubble Space Telescope (FUV - blue visible), the Chandra X-ray Observatory, and the XMM-Newton. An open source database is released that contains panchromatic stellar energy distributions (hereafter SEDs) in the range of 5$\times10^{-4}$-5.5 $\mu$m.\footnote{Database: \href{https://archive.stsci.edu/prepds/muscles/}{https://archive.stsci.edu/prepds/muscles}} Due to ISM absorption, the Ly-$\alpha$ emission line is reconstructed using a model that fits the line wings (\citeauthor{Youngblood2016} \citeyear{Youngblood2016}), and subsequently the EUV regime is created using an empirical scaling relation based on the Ly-$\alpha$ flux (\citeauthor{Linsky2014} \citeyear{Linsky2014}). The visible- and IR-regimes for each SED are synthetic photospheric spectra from the PHOENIX atmosphere models (\citeauthor{Husser2013} \citeyear{Husser2013}). The complete SEDs of all stars considered in this study are shown in figure \ref{fig:spectra}. These quiescent spectra are used in the chemical kinetics code to create an equilibrium profile before any flare event occurs.  

\subsubsection{Synthetic Flare Spectra Implementation}

We simulate stochastic flaring of the host star using the fiducial flare program developed by \citet{Loyd2018b}\footnote{Github: \href{https://github.com/parkus/fiducial_flare}{parkus/fiducial\_flare}} so that we can evaluate the effects of flaring on timescales longer than the existing UV observations of the host stars. 
The fiducial flare spectrum is based on an energy budget from actual flare measurements. After obtaining the fiducial flare spectrum, they are normalized to the Si IV doublet of the observed spectrum values from the MUSCLES-spectra, as seen in the box-figure in fig \ref{fig:spectra}. Finally, the quiescent spectra are added to the synthetic spectra such that the included flares always have energies larger and/or equal to quiescent values. 

The model also includes the option to produce a light-curve containing a series of flare events. Flare frequencies are derived from the \textit{flare frequency distribution} (hereafter FFD), which is assumed to follow a power-law model (\citeauthor{Hawley2014} \citeyear{Hawley2014};  \citeauthor{Loyd2018a} \citeyear{Loyd2018a}, \citeyear{Loyd2018b}; \citeauthor{Davenport2014} \citeyear{Davenport2014}; \citeauthor{Ilin2018} \citeyear{Ilin2018}). To be more specific, \citet{Loyd2018b} used time-resolved spectral data of M-dwarfs to fit the general power-law equation,
\begin{equation}
\label{eq:FFD}
    \nu = 10^{\beta}\left(\frac{\delta}{\delta_{\mathrm{ref}}}\right)^{\alpha}
\end{equation}
where $\nu$ is the occurrence rate (per day) of a given flare, $\delta$ is the equivalent duration of the flare, and $\delta_{\mathrm{ref}}$ is the reference duration value (set equal to $\delta_{\mathrm{ref}} = 1000$ sec). The equivalent duration is defined as the fractional flux, between observed flare and quiescent spectra, integrated over time. In mathematical terms,
\begin{equation}
    \label{eq:eqdur}
    \delta = \int_{t_0}^{t_e} \frac{F_f-F_q}{F_q} \mathrm{d}t
\end{equation}
where $F_f$ and $F_q$ are the quiescent and flare flux integrated over a specific waveband, and $t_0$ and $t_e$ are the start and ending times of the flare-event. A more detailed description and visualisation of both equations can be found in \citet{Hawley2014} and \citet{Loyd2018b}. Finally, $\alpha$ and $\beta$ in eq. \ref{eq:FFD} are fixed parameters found by fitting flare data to power law models. The values of $\alpha$ and $\beta$ depend on the type of M-dwarfs used to fit the parameters. The appropriate values for the stars in this study are $\alpha = -0.77$ and $\beta = 0.45$ (\citeauthor{Loyd2018b} \citeyear{Loyd2018b}).

\begin{figure}
    \centering
    \includegraphics[scale=0.54]{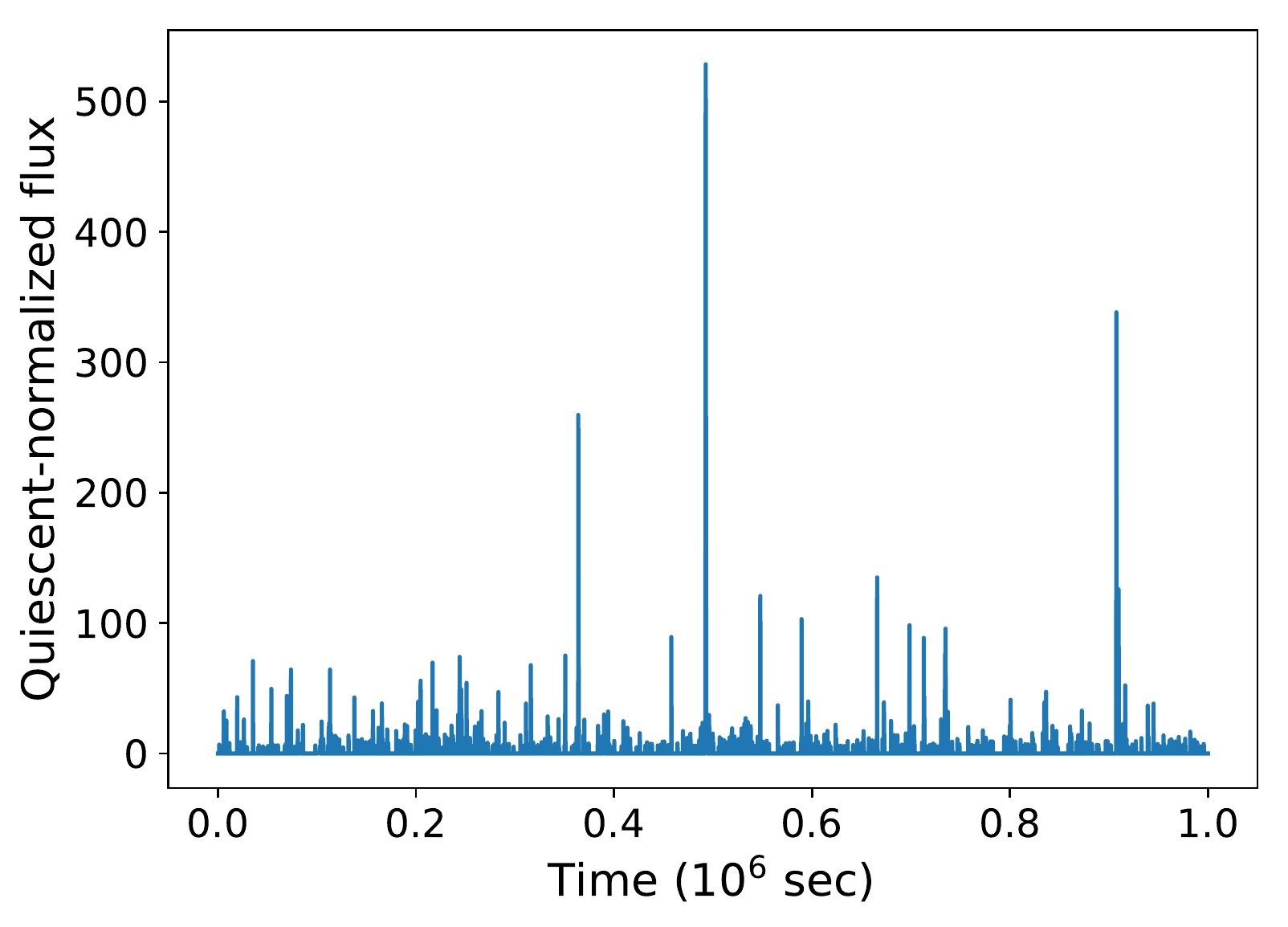}
    \caption{ Light curve from the fiducial flare program. This lightcurve has been created assuming an inactive M-dwarf with a flaring blackbody temperature of 9000 K. A total of 515 flares are shown within this lightcurve.}
    \label{fig:lightcurve}
\end{figure}

The final light-curve used is shown in figure \ref{fig:lightcurve}. This light-curve is created by integration over the Si IV doublet. The flare events are distributed following the Poisson distribution. Note that the light-curve is normalised to the quiescent flux values such that integrating over a single flare equals the equivalent duration of that particular flare (see eq. \ref{eq:eqdur}). Re-writing the inner-integral term in eq. \ref{eq:eqdur} gives the time-dependent flare spectra,
\begin{equation}
    F_f(t) = F_q \cdot (a(t) + 1)
\end{equation}

where $a(t)$ is the quiescent-normalized flux. Where \cite{Chen2021} used this multiplication factor for the UV-regime of the quiescent spectrum, we will use it  as a time-dependent multiplication factor to the synthetic flare spectrum (see figure \ref{fig:spectra}). When running the chemical kinetics code, the light-curve is initialized beforehand at fixed time steps. During integration, the stellar
spectrum can then be treated as a time-dependent variable by using linear interpolation. The step sizes, $dt$, during integration are regulated by the estimated truncation error. As time evolves, this error will become smaller and thus the step sizes increase. To ensure that no flare spectrum is missed during integration, these variable step sizes are limited to the maximum amount of time between the current time step $t$ and the time at the start of a new flare event. Due to a long integration time we only integrate for 1e6 seconds (see fig. \ref{fig:lightcurve}), i.e. approx. 11 days.

\subsection{Chemical Model}
We use the open source chemical kinetics code VULCAN\footnote{Github: \href{https://github.com/exoclime/VULCAN}{exoclime/VULCAN}} (\citeauthor{vulcan2017} \citeyear{vulcan2017}; \citeyear{Tsai2021}) for modelling the chemical abundances of the atmospheres.
 This code is able to solve a set of one-dimensional mass continuity equations, which are time dependent. The elements included are nitrogen, carbon, hydrogen, oxygen, and helium which make up 74 different species in total. The chemical reactions used are the default validated photo-chemical network of VULCAN. It consists in total of 948 reactions, that includes the forward- and backward reactions and photo-dissociation reactions. As input parameters, VULCAN requires planetary- and stellar characteristics, a temperature profile, and a stellar spectrum, which are discussed more thoroughly in the next sections. The initial molecular abundances are calculated using the equilibrium chemistry code FastChem\footnote{Github: \href{https://github.com/exoclime/FastChem}{exoclime/FastChem}} (\citeauthor{Stock2018} \citeyear{Stock2018}), which is coupled to VULCAN. FastChem is an analytical computer program that makes use of the solutions of the coupled non-linear equations of law of mass action and the element conservation to calculate chemical equilibrium.

\subsection{Atmospheric model}

\subsubsection{Compositions}

Two out of three planets considered in this study are on the brink of being super-Earths or sub-Neptunes. Also taking in to account that the radii of these planets have not been observed yet, we investigate two type of exoplanets, gaseous and rocky worlds. For the elemental abundances we make use of two specific atmospheres as a clear distinction between the two cases. For the gas giants we use solar abundance (hereafter H$_2$-dominated atmospheres) and for the rocky worlds we use an Earth-like elemental abundance (hereafter N$_2$-dominated atmospheres), i.e. $\mathrm{O}/\mathrm{H} \approx 39.99$; $\mathrm{C}/\mathrm{H} \approx 0.04$; $\mathrm{N}/\mathrm{H} \approx 155.96$; and $\mathrm{He}/\mathrm{H} \approx 0.001$. These elemental abundances are fed into the chemical equilibrium code FastChem to obtain the initial abundance for VULCAN.\footnote{This particular coupled-option between the chemical equilibrium and chemical kinetics code is already implemented in the latest version of VULCAN}

\subsubsection{Temperature profiles}
\begin{figure*}
    \hspace{2em}
    \centering
    \includegraphics[scale=0.86]{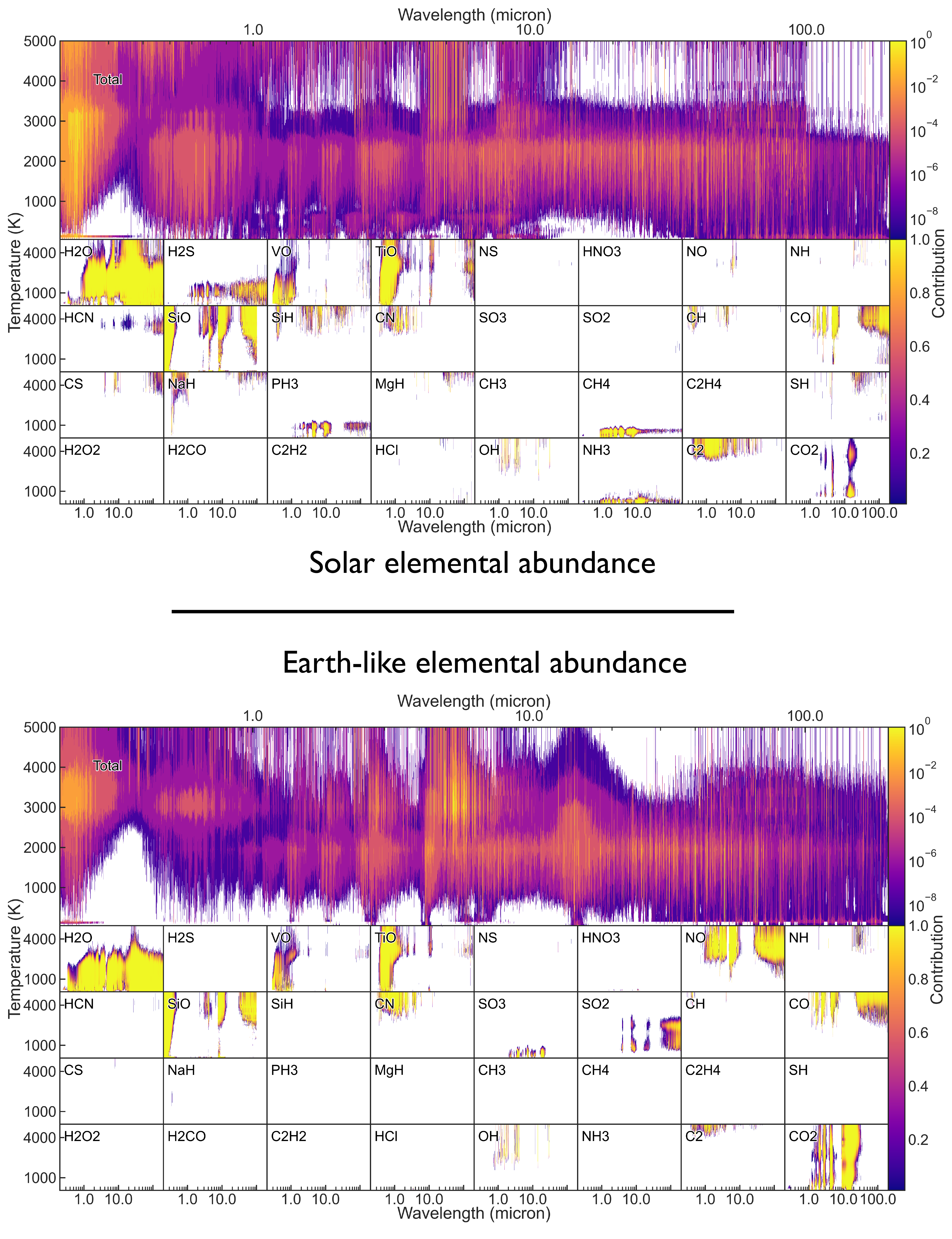}
    \caption{Contribution of each species (small boxes) to the total opacity (top subpanel) with Solar elemental abundance (top) and Earth-like elemental abundance (bottom). All contributions are shown for $P = 10^{-4}$ bar. In reality, these contribution plots exist for all pressure levels, though we only show one as an example. For each species the vertical axis represents the temperature and the horizontal axis represents the wavelength. For the individual species, the color represents the contribution to the total opacity. The top panel of both figures represents the total contribution of all species combined and can be used to determine which species contribute the most at a certain temperature and wavelength.}
    \label{fig:opac_fullgrid}
\end{figure*}

The temperature profiles of all planets in this paper are calculated using the radiative transfer code HELIOS\footnote{Github: \href{https://github.com/exoclime/HELIOS}{exoclime/HELIOS}} (\citeauthor{Malik2017} \citeyear{Malik2017}; \citeauthor{Malik2019} \citeyear{Malik2019}). This code numerically solves the radiative transfer equation using the two-stream approximation. The solutions to these equations are then used to iteratively solve for the temperature-pressure profile. In order to use this code for the temperature profiles, it requires planet system parameters (table \ref{tab:planet_params}), the stellar spectrum, and the opacity tables. 
In this study we assume that the temperature-profiles are not influenced by the stellar flares and are thus static in time.
The TP-profile is therefore calculated only once using blackbody spectra of each stellar-system, which is then used as input for VULCAN and remains constant throughout the simulation. The opacity tables are created using the line-opacities from the open source opacity database\footnote{Database: \href{https://dev.opacity.iterativ.ch/\#/}{opacity.world}} (\citeauthor{Grimm2015} \citeyear{Grimm2015}). The k-tables are sampled for a wavelength range of 0.02 $\mu$m up until 200 $\mu$m, with a resolution of $\triangle \lambda/\lambda = 1000$. 
The included absorbers in the atmospheres are carefully selected by considering only the most affecting and abundant species in the atmospheres. For both the H$_2$- and N$_2$-dominated atmospheres, opacity tables have been created that includes 30+ species on a TP-grid that ranges from $50 \mathrm{\ K} \leq T \leq 6000 \mathrm{\ K}$ and $10^{-6}\mathrm{\ bar} \leq P \leq 10^3 \mathrm{\ bar}$ to investigate what molecular species have most contribution to the total opacity. A special opacity grid is created for all species, wavelengths, temperatures, and selected pressures (see figure \ref{fig:opac_fullgrid}). In these plots, the color represents the contribution of the opacity of a certain species to the total opacity as a function of temperature and wavelength at a certain pressure level. These grids are therefore an intuitive tool to determine which species have most impact in these radiative transfer computation for different types of planets.  As an illustration, figure \ref{fig:opac_fullgrid} shows the opacity grid for $P =  10^{-4}$ bar.
The final species included when creating the TP-profile are then selected by only considering species that would be most relevant for that type of atmosphere by looking at the expected pressures and temperatures. For the H$_2$-dominated atmospheres, the included species are CH$_4$, CN, CO, CO$_2$, H$_2$O, H$_2$S, NH$_3$, NaH, PH$_3$, SiO, TiO, and VO, while for the Nitrogen-dominated planets we included H$_2$O, CO$_2$, SiO, TiO, SO$_3$, SO$_2$, and VO. 

\begin{figure*}
\hspace{-3 em}
    \centering
    \includegraphics[scale=0.63]{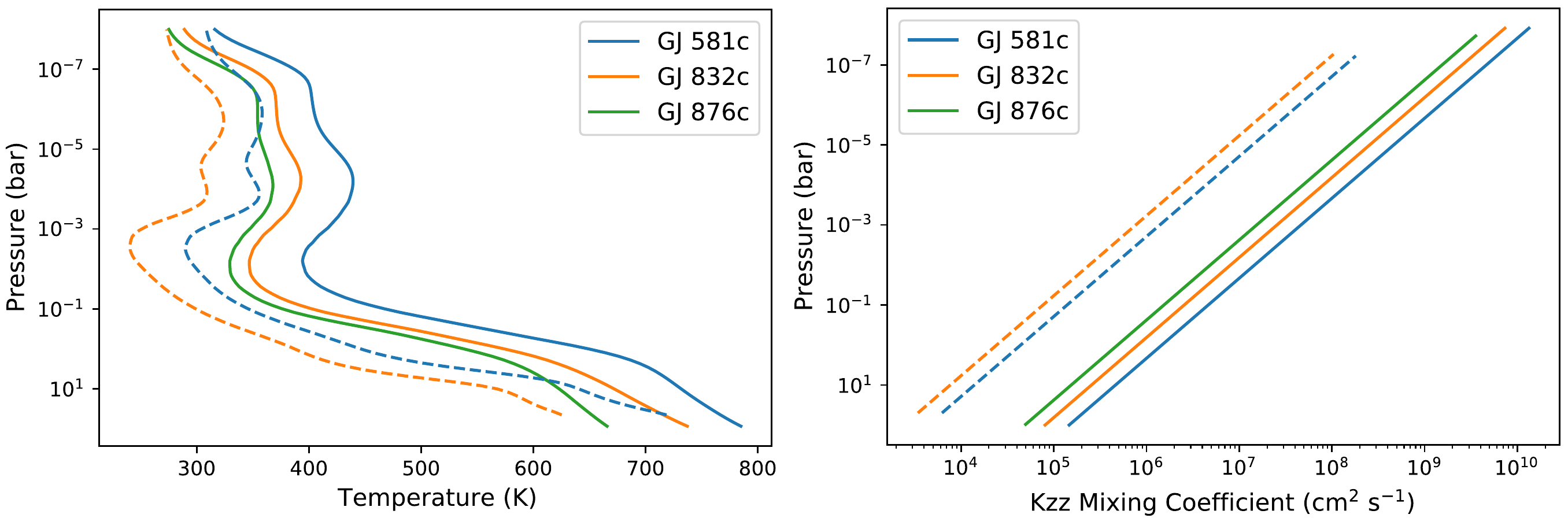}
    \caption{Temperature profiles (left) created with HELIOS for H$_2$-dominated atmospheres (solid lines) and N$_2$-dominated atmospheres (striped lines) and vertical mixing profiles (right).}
    \label{fig:input_params}
\end{figure*}

The final temperature profile for each individual planet is shown in figure \ref{fig:input_params}. As we deal with two planets that can be estimated as gaseous as well as rocky worlds, we get a total of 5 case studies.

\subsubsection{Vertical Mixing}
\label{tab:vertical_mixing}
Mixing in the atmosphere is considered only in the vertical direction by means of eddy- and molecular diffusion.
In general, eddy diffusion is most dominant in the lower and intermediate parts of the atmosphere, where the species are well mixed. The eddy diffusion coefficient, $K_{zz}$, is considered as a free parameter in this study. In a one-dimensional model it is common to determine its value using the mixing-length formulation (i.e. $K_{zz} = C l w_t$, with $C$ being a constant, $l$ the mixing length, and $w_t$ the turbulent velocity), where free-convection is assumed in this model. It remains a difficult task to solve for the diffusion coefficient. Various models simulating H$_2$-dominated atmospheres adopt values between $K_{zz} = 10^8$ - $10^{12} \mathrm{\ cm^2 \ s^{-1}}$ (\citeauthor{Parmentier2013} \citeyear{Parmentier2013}; \citeauthor{Moses2012} \citeyear{Moses2012}; \citeauthor{Miguel2014} \citeyear{Miguel2014}).

In this study, we adapt the expression for the eddy diffusion $K_{zz}$ from \citeauthor{Moses2021} \citeyear{Moses2021},
\begin{equation}
    K_{zz} = 5\times 10^{8}[P(\mathrm{bar})]^{-0.5}\left(\frac{H_{\mathrm{1 \ mbar}}}{620 \mathrm{\ km}}\right)\left(\frac{T_{\mathrm{eff}}}{1450\mathrm{\ K}}\right)^4
    \label{eq:kzz}
\end{equation}
Where $P$ is the pressure profile of the atmosphere in bars, $H_{1 \mathrm{\ mbar}}$ is the scale height at 1 mbar and T$_{\mathrm{eff}}$ is the effective temperature. The exact profile of each of the case studies can be seen in figure \ref{fig:input_params}.

Going to the upper atmospheres the most dominant mixing method is molecular diffusion. The molecular diffusion coefficient, $D_{zz}$, is determined for either a hydrogen or nitrogen dominated atmosphere, depending on the most abundant species present in the atmosphere. We use a nitrogen, N$_2$, dominated atmosphere for the smaller rocky-type planets and a hydrogen, H$_2$, dominated atmosphere for the gaseous planets. 
\subsubsection{Boundary conditions and atmospheric escape}

In this study we include upper boundary conditions with an outward particle flux and we use a closed lower boundary with no upward and/or downward particle flux. We simulate the N$_2$-dominated planets as rocky worlds with a solid surface as lower boundary condition when creating the TP-profiles, which introduces some surface heating. The upper boundary condition has an upward flux due to atmospheric escape of particles.  Escape of particles at the top of the atmosphere can occur due to (non-)thermal escape processes, such as hydrodynamic escape due to XUV-heating, sputtering and charge exchange escape due to stellar winds, or photochemical escape, to name a few. These escape mechanism all work separate from each other, though it is certainly possible that various escape processes operate simultaneously. The escape of lighter particles due to these escape processes, however, can be limited by the available amount at the top of the atmosphere. Diffusion at the homopause can supply the light particles for these escape processes, but also puts a limit on the final amount that is able to escape. This process is called \textit{diffusion limited escape}. In this study we therefore assume that the escape is restricted to the diffusion at the homopause \citet{Tsai2021},

\begin{equation}
    \Phi_{DL} = D_t n_i \cdot \left(\frac{1}{H_t} - \frac{m_i}{N_{\mathrm{avo}}k_B T}\right)
    \label{eq:difflimesc}
\end{equation}
where $\Phi_{DL}$ is the diffusion limited flux, $D_t$ is the molecular diffusion coefficient, $n_i$ is the number density of species $i$, $H_t$ is the scale height at the top of the atmosphere, and $T$ is the temperature, all evaluated at the top of the atmosphere. The constants $m_i$, $N_{\mathrm{avo}}$, and $k_B$ represent the molecular weight of species $i$, Avogadro's constant, and Boltzmann's constant respectively. Note that the diffusion limited flux is directly proportional to the number density. 

\subsection{Emission and Transmission Spectra}
The spectral characterization of the studied atmospheres is calculated using the open-source, radiative transfer Python package \textit{petitRADTrans}\footnote{Documentation: \href{https://petitradtrans.readthedocs.io/en/latest}{petitradtrans.readthedocs.io}} (\citeauthor{Molliere2019} \citeyear{Molliere2019}, \citeyear{Molliere2020}). The opacities of the most abundant molecules and atoms that make up the photo-chemical network and fluctuate most throughout the simulation are included, i.e. for the gaseous planets: H$_2$O, CO$_2$, CO, CH$_4$, NH$_3$, OH, HCN, C$_2$H$_2$, and for the rocky worlds: H$_2$O, CO$_2$, CO, OH, and O$_3$. All cases also include collision induced absorption (CIA) of H$_2$-H$_2$ and H$_2$-He, and for the nitrogen dominated worlds also N$_2$-N$_2$. 
 All opacities included have low resolution spectra  of $\lambda/\triangle\lambda = 1000$ for the calculation of the emission and transmission spectra. 

\section{Results}
\label{sec:results}

In this section we present the results which can be subcategorized in two parts. We first present the evolution of the molecular and atomic abundances.  Secondly, we show the evolution of the emission and transmission spectra. 

\subsection{Chemical Compositions}

\subsubsection{H$_2$-dominated atmospheres}

\begin{figure*}
 \hspace{-2em}
    \centering
    \includegraphics[scale=0.95]{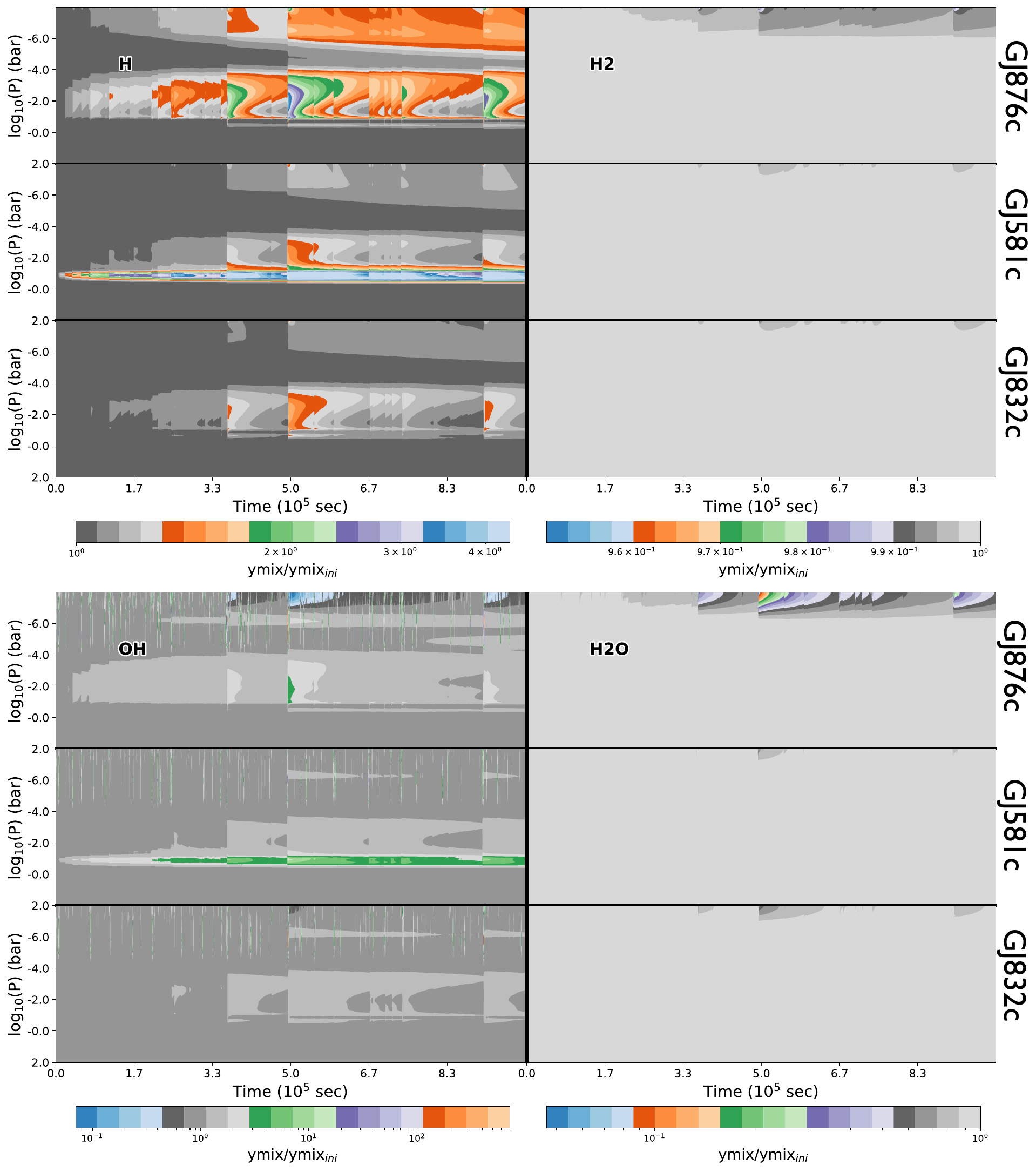}
    \caption{Evolution of the normalized abundance of atomic hydrogen (upper left), molecular hydrogen (upper right), OH (lower left), and H$_2$O (lower right),  for GJ 876c (upper panels), GJ 581c (middle panels), and GJ 832c (lower panels) of hydrogen dominated atmospheres. The color represents the change in abundance normalized to the initial composition. }
    \label{fig:Jov_abund_change_A}
\end{figure*}

\begin{figure*}
\hspace{-2em}
    \centering
    \includegraphics[scale=0.93]{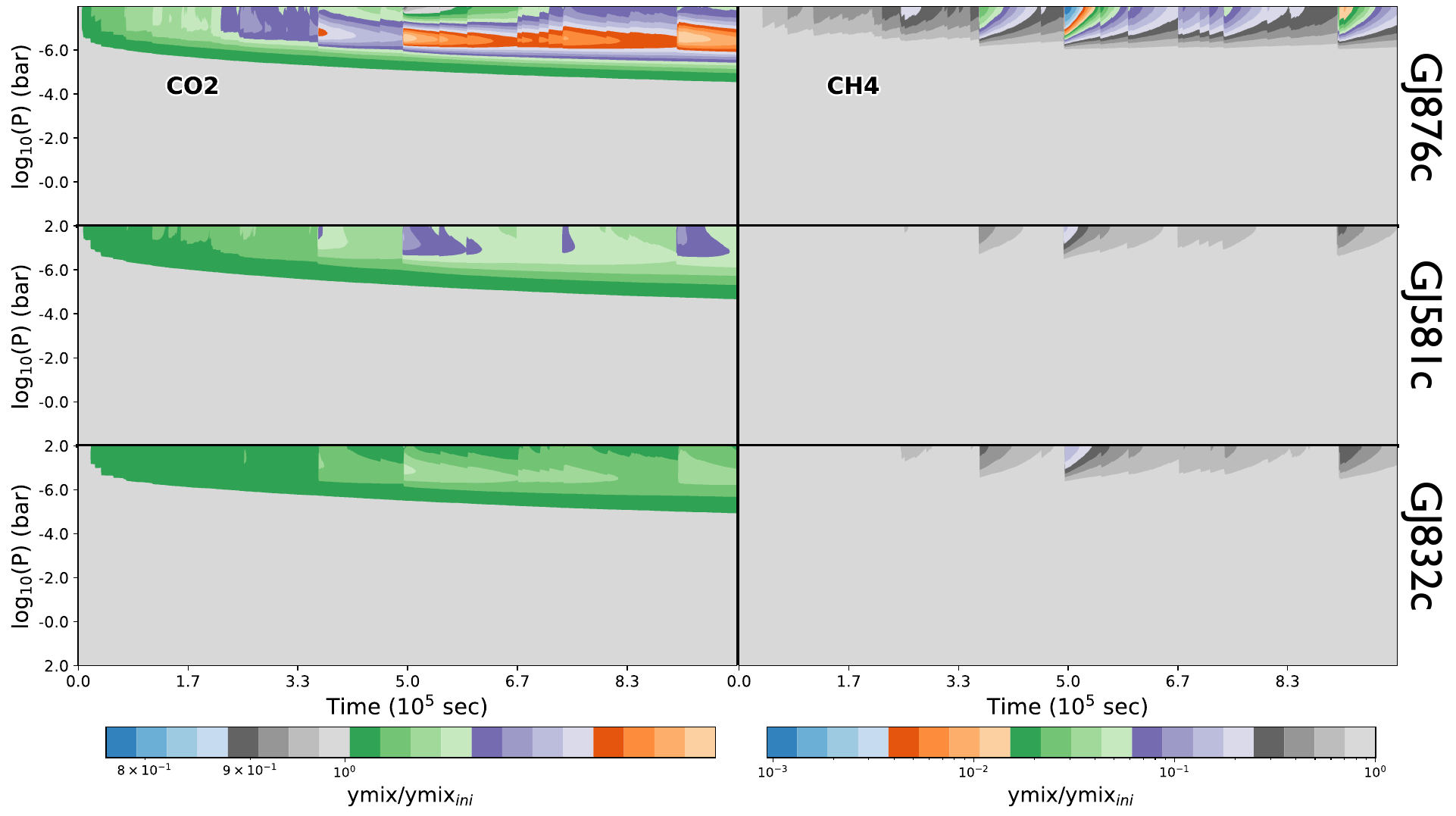}
    \caption{Evolution of the normalized abundance of CO$_2$ and CH$_4$ for the hydrogen dominated atmospheres, similarly as in figure \ref{fig:Jov_abund_change_A}. }
    \label{fig:Jov_abund_change_C}
\end{figure*}

The changing normalized mixing ratios of various species included in the photo-chemical network are shown in figures \ref{fig:Jov_abund_change_A} and \ref{fig:Jov_abund_change_C} for the H$_2$-dominated atmospheres. Overall, each species shows a general trend where an increase or decrease is found after a flare event. The variations in molecular and atomic abundances strongly depend on the sizes of the flares. As expected, large flares have significantly more impact on the abundance change than the smaller flares. Nonetheless, some species show a gradual change in abundance, indicating that these changes accumulate throughout time due to the recurring, smaller flares. The effect of single flare events and recurring flare events are different for each species. Noticeable is that the gas giant, GJ 876c, reveals most changes in composition throughout the atmosphere compared to the smaller planets GJ 581c and GJ 832c. The most stable atmosphere can be found on GJ 832c, which is most likely due to its small size and large orbit radius relative to the other planets.

\subsubsection*{Single big flare events}

Species that show abrupt changes and recovery in abundance, as a result of single flare events, are H, OH, H$_2$O, and CH$_4$. In figures \ref{fig:Jov_abund_change_A} - \ref{fig:Jov_abund_change_C} these flare events can be recognized by the vertical, noisy lines. For \textbf{hydroxide} each flare event causes an increase in abundance, after which it quickly recovers its value before a new flare event occurs. The change in abundance can at some points even increase to up to 2 orders in magnitude. The bigger flare events are recognized by a decrease in abundance at the top of the atmosphere. \textbf{Atomic hydrogen}, on the other hand, seems to recover slower and therefore has a smoother evolution in abundance change. The abrupt changes can be seen mostly in the intermediate atmosphere around $10^{-3.6} < P < 10^{0.8}$ bar. Similar to the change in hydrogen abundance, \textbf{water} shows a slower recovery of the abundance changes due to the flare events, causing a permanent change in abundance in the upper atmosphere ($P <10^{-5.8}$ bar), with some fluctuations due to the larger flare events. These changes are mostly found on GJ 876c, and remain imperceptible on GJ 581c and GJ 832c. Finally, \textbf{methane} shows more fluctuations in the upper atmosphere due to flare events and is therefore more sensitive to the stellar flux changes. On GJ 876c these fluctuations can decrease to up to 3 orders in magnitude. 

\subsubsection*{Cumulative changes}

Species that show accumulative change in abundance are H, H$_2$, OH, and CO$_2$. For molecular and atomic \textbf{hydrogen} it can be seen that the gradual change in abundances mostly occur in the upper atmosphere. The escape of particles in this region is the cause of this accumulating change. More interestingly is the seemingly permanent abundance change in atomic hydrogen mid-atmosphere (1-10$^{-2}$ bar) for GJ 581c. A possible explanation could be that the combination of vertical mixing and flaring events caused the composition to find a new equilibrium in this region. This long-lasting change in hydrogen increases to up to almost one order in magnitude as compared with the initial abundance. Conversely, the gradual change of \textbf{hydroxide} can be seen best on GJ 876c and GJ 581c around $P > 10^{-5.8}$ bar. For hydroxide there also seems to be a permanent change mid-atmosphere (1-10$^{-2}$ bar) on GJ 581c. This permanent abundance change seems to be an order of magnitude bigger than initial abundances. Finally, the accumulative change for \textbf{carbon dioxide} appears around $P \sim 10^{-5.8}$ bar, though the changes are minimal. The changes in abundances do not only accumulate throughout time but also seem to go deeper in the atmosphere over time. These gradual changes, however, are small and only differ a factor $< 10$ with the initial abundances.\\
\\
 Even though the trends in the abundance plots look similar for all three planets, there are some key differences worth noting. Firstly, the biggest and coldest planet, GJ 876 c, seems least stable of all three planets considered in this study. This is most likely due to it's higher abundance in hydrogen due to its relatively large size. 
  
\begin{figure*}
 \hspace{-2em}
    \centering
    \includegraphics[scale=0.64]{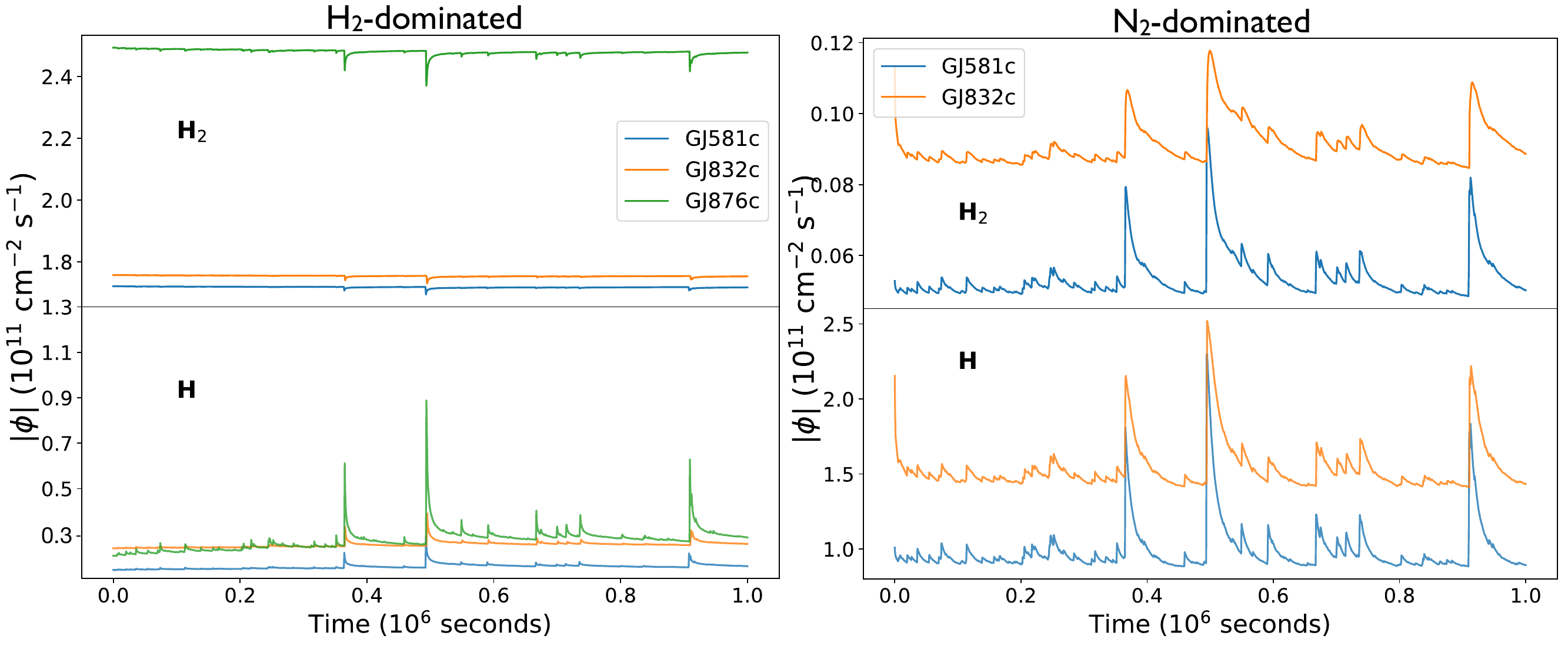}
    \caption{Absolute escape rates of atomic (bottom panels) and molecular (top panels) hydrogen for H$_2$-dominated (left figures) and N$_2$-dominated (right figures) atmospheres as a function of time.}
    \label{fig:escape_rates}
\end{figure*}

  When looking at the escape rates of molecular and atomic hydrogen of the gas giants (figure \ref{fig:escape_rates}) it can be seen that the escape rate of molecular hydrogen is significantly higher on GJ 876c than the other two cases. The number density of molecular hydrogen is therefore highest on GJ 876c and we expect to see biggest changes in abundance in this atmosphere as well (see figure \ref{fig:Jov_abund_change_A}). For all three cases the escape rate seems quite constant over time, with the exception of a few large flare events. The abundance change in molecular and atomic hydrogen should therefore also be quite stable in the upper atmosphere, as can be seen in figure \ref{fig:Jov_abund_change_A}. In more conceivable numbers, the average mass loss rate on GJ 876c is $\sim 6.8\times 10^{-22} \mathrm{\ M_p \ s^{-1}}$ (with $\mathrm{M_p}$ the planet mass). We divide by the total planet mass such that we can make a better comparison with other known mass loss rates. The mass loss rate found here can be thought of as insignificant loss to the total mass in short periods of time. As a comparison, we cite here the example of the hot sub-Neptune GJ 436b, that has a measured mass loss rate of $\sim 7.6 \times 10^{-21} \mathrm{\ M_p \ s^{-1}}$ (\citeauthor{Ehrenreich2015} \citeyear{Ehrenreich2015}). Even though GJ 436b has a lower mass ($\sim$22.1$M_{\oplus}$), the relative mass loss rate is a factor 10 higher than the relative mass loss rate of GJ 876c. This is because GJ 436b has a small orbital period around its host star and therefore falls within the hydrodynamic escape regime due to high irradiation, as opposed to the diffusion limited escape regime.\\
  \\ 

Looking at all H$_2$-dominated atmospheres, the changes in abundances for most species occur in the mid- and upper atmosphere, going up to at most $\sim 1$ bar (e.g. OH for GJ 581c), in agreement with results by \cite{Venot2016}. The flare event from AD Leo is, however, smaller than the biggest flare considered in this study when looking at the relative differences in  peak fluxes. The changes in abundances therefore differ slightly as well. One main difference between their results and the results presented here is the change in hydrogen in the upper atmosphere. In \cite{Venot2016} the abundance for atomic hydrogen remains the same, whereas we certainly see a change in atomic hydrogen over time. 

The inclusion of repeated flaring and atmospheric escape therefore seem to have a great impact on the hydrogen abundances in this region. In addition, the planets they considered have a higher effective temperature (with a difference of $\sim 100$ K compared to the effective temperatures used here). Atomic hydrogen is therefore the dominant species in their upper atmospheres instead of molecular hydrogen. The dissociation rate of molecules in to atomic hydrogen is thus slightly lower in those atmospheres, which might explain the difference in abundance change.

In addition to the single flare event from AD Leo, \cite{Venot2016} also looked at repeated flaring by re-using the same flare event. In total they simulated the atmospheres for $\sim 10^6$ seconds. They found that the species show a gradual change in abundance - a trend that can be found in this study as well. This might indicate that the accumulating changes are actually not caused by the escape of particles in the upper atmospheres, but rather by the recurring flares, as they did not include atmospheric escape in their calculations. Furthermore, the gradual change found in \cite{Venot2016} seemed to converge to limiting values. Whereas in this study we find that for some species the accumulating changes have not converged yet to specific values (e.g. H, H$_2$, and H$_2$O). Including minor flares thus seem to have an impact on the abundances of species over time.

\subsubsection{N$_2$-dominated atmospheres}

\begin{figure*}
    \centering
    \includegraphics[scale=0.9
    ]{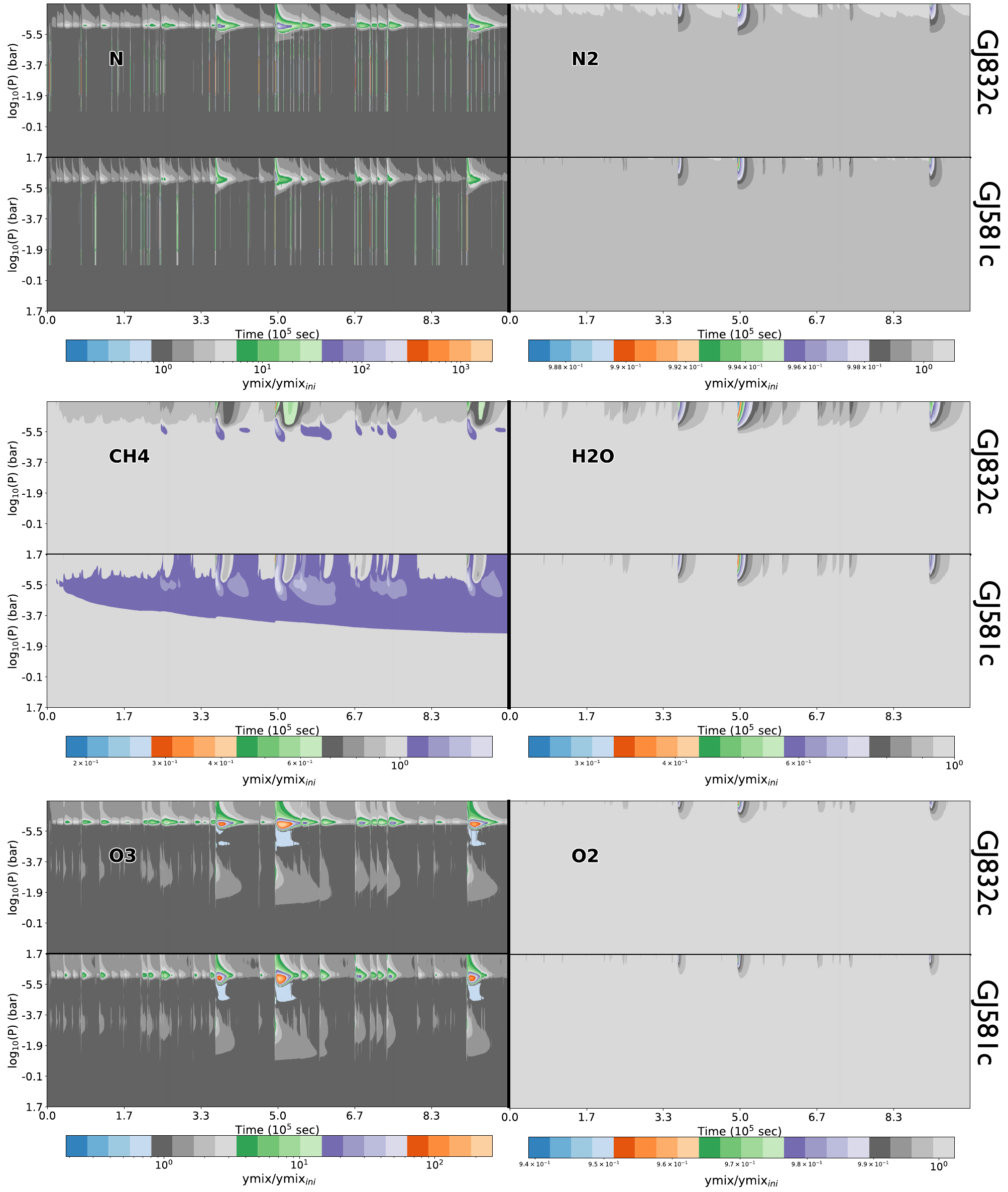}
    \caption{Evolution of the normalized abundance of N, N$_2$, and the bio-signatures CH$_4$, H$_2$O, O$_3$, and O$_2$ or GJ 832c (upper panels), and GJ 581c (lower panels) of nitrogen dominated atmospheres. The color represents the change in abundance normalized to the initial composition. }
    \label{fig:abundance_change_RW}
\end{figure*}

Figure \ref{fig:abundance_change_RW} represents the abundance change plots for the two rocky world cases of GJ 832c and GJ 581c. Overall, most species show again the general flaring trend. For some species we see, however, a gradual change in their composition. Similar to previous results, the effect of the individual flares and the recurring flares differ for each species. Additionally, we see that the atmospheres of GJ 581c and GJ 832c react very similar, besides the change in abundance of methane. A critical difference  between these planets is their semi-major axes, which results in different temperature-pressure profiles (fig. \ref{fig:input_params}) that affect the overall methane abundances.

\subsubsection*{Single big flare events}

The impact of the individual flare events in abundance can be seen for the species N, CH$_4$, H$_2$O, O$_3$, O$_2$, and N$_2$. First of all, for \textbf{atomic nitrogen}, the individual flares can be identified as the vertical, abrupt lines in the intermediate atmosphere ($10^{-5.5} \mathrm{\ bar} \leq P \leq 10^{-0.1} \mathrm{\ bar}$). These lines present a rapid response of the abundance to the change in UV radiation from the host star. The recovery of the abundance to the initial values after the flare has passed is quite immediate as well. Also in the upper atmosphere ($P < 10^{-6}$ bar) the effect of the individual flares can be seen, though the recovery is slower, which again creates a smoother profile. \textbf{Ozone} shows a similar evolution in abundance, with abrupt changes and recovery in the intermediate atmosphere and smoother transitions between flares in the upper atmosphere. For \textbf{methane}, on the other hand, the individual peaks of the flares can only be seen in the upper atmosphere. The recovery is also relatively slow compared with the time intervals between the flares, and thus it never fully recovers to it's initial abundance, as seen on GJ 832c. For GJ 581c this feature is absent in the upper atmosphere. Similarly, \textbf{water} shows a decreases in abundance in the upper atmosphere in response to the flare events. Larger flares do seem to have an even bigger impact on the abundance, sometimes decreasing the abundance with a factor of up to 5. \textbf{Oxygen} seems quite resistant to the smaller flare events. The larger flare events are quite distinctive and show a very smooth recovery afterwards. A decrease of at most $\sim 2.5\%$ is found due to flare events in the upper atmosphere (> 100 km above surface level). Considering that the mixing ratio is high throughout the atmosphere ($\sim 0.2$), this oxygen depletion is substantial. At last, for \textbf{molecular nitrogen} we see the distinct larger flare events in the upper atmosphere ($P < 10^{-5.5}$ bar). Recovery after large flare events go quite gradually.

\subsubsection*{Cumulative change}

Species that show accumulative changes over time are CH$_4$, O$_3$, and N$_2$.  For \textbf{molecular nitrogen} this small, but present, feature occurs in the upper atmosphere around $P < 10^{-5.5}$ bar on GJ 832c. The changes in these regions are, however, insignificant. \textbf{Methane} shows the accumulative change in increasing abundance at $P < 10^{-3.7}$ bar. These gradual changes are relatively small as well and can be neglected.
Finally, the \textbf{ozone} abundance seems to change gradually in the upper atmosphere as well, around $P \sim 10^{-5.5}$ bar. This abundance change is quite substantial compared with the previous species. After 11 days the abundance differs an order in magnitude compared with the initial abundance. In addition, the intermediate atmosphere appears to have a small permanent change in abundance.  
 All in all, it seems like the inclusion of smaller flares leads to these gradual changes in the N$_2$-dominated atmospheres.\\
\\

The escape rate of atomic and molecular hydrogen is similar in this case as compared with the H$_2$-dominated atmospheres. For both atomic and molecular hydrogen the escape rate seems to decrease over time (see figure \ref{fig:escape_rates}) on GJ 832c. The lack of hydrogen in these atmospheres ensures that there is not enough atomic hydrogen produced due to photo-dissociation, and thus the hydrogen supply keeps decreasing. The escape rate of hydrogen on both planets seems to converge to smaller values. 
Comparing the mass loss rate with known rates on Earth shows some differences. Particle escape on Earth is driven by two main mechanisms, thermal Jeans escape and the non-thermal charge exchange escape. The total mass outflow due to these escape processes is approximately $3.4\times 10^{-25} \mathrm{\ M_p \ s^{-1}}$ (\citeauthor{gronoff2020} \citeyear{gronoff2020}). On GJ 832c this mass rate is significantly bigger with an average value of $\sim 1.2\times 10^{-22} \mathrm{\ M_p \ s^{-1}}$, a factor $\sim 350$ times greater than on Earth. One reason is the higher temperatures on GJ 832c causing the atmosphere to be more prone to thermal escape mechanisms. Additionally, we do not include condensation in this study, enforcing all species to be in gas state, and thus making hydrogen more available due to processes such as  photo-chemistry. 

\citet{Tilley2018} studied the evolution of Earth-like planet atmospheres exposed to flares, including proton impacts, of an M star (AD Leo). As opposed to this study, the atmospheres modelled in \citet{Tilley2018} are based on the knowledge of Earth's biological and interior input to the atmosphere. In our study, we remain agnostic to the possibility of life on these worlds and the impact that potential volcanoes and life might have on the atmospheres. Comparing this study with \citet{Tilley2018} thus gives us more insight in identifying differences for future planetary searches and characterisation.
In general these differences can be found in several species, such as ozone, oxygen, methane and water abundances. Important to note is that the upper atmosphere of \citet{Tilley2018} is at $\sim$ 65 km, while in this study the upper atmosphere extends to $\sim$ 120 km. We therefore focus on the differences and similarities around $P \approx 10^{-3}$ bar for both planets. In \citet{Tilley2018}, it was found that electromagnetic-only (hereafter EM-only) flaring changes \textbf{ozone} abundances insignificantly at 65 km above surface level, but it did cause depletion of ozone in their upper atmosphere. In contrast, in the results presented here, the ozone abundances show, in general, an increasing trend with exception of the abrupt decreases at the larger flare events. At $P \approx 10^{-3}$ bar there is a maximum increase of a factor of $\sim 6$ in ozone abundance. In this study, ozone is produced through the reaction between atomic and molecular hydrogen. The production and loss of oxygen is regulated mainly by H-bearing species, and the same holds for ozone. When comparing water abundances, we find that for altitudes higher than 10 km, the atmosphere in \citet{Tilley2018} is dryer compared to the atmospheres in this study (mixing ratios of $\sim 10^{-4}$ and $\sim 10^{-3}$ resp.). This extra amount of water dissociates into lighter molecules to eventually form ozone.
\textbf{Oxygen} abundances do not show a significant change in GJ 832c and GJ 581c, nor does it change significantly when compared with previous studies. Comparing the \textbf{oxygen} abundances with the ones found in \citet{Tilley2018} shows some similarities in the lower atmosphere (< 100 km above surface level).
The changes found in the atmosphere for \textbf{methane} are a $\sim 0.8$\% decrease in mixing ratio after big flare events on GJ 832c and a $\sim$2.9\% increase in mixing ratio on GJ 581c at 65 km above surface level. Also taking in to account that the mixing ratios of methane at this pressure level is $\sim 10^{-18}$ and $\sim 10^{-21}$ on GJ 832c and GJ 581c respectively, we find that the changes in methane due to EM-flaring is most likely not observable.  
In \citet{Tilley2018} they found that the change in methane over a time-span of 10 years is significant. 
This difference in results on methane abundance is because we do not include any biological sources that produce CH$_4$, as opposed to the results in \citet{Tilley2018} 
Finally, \textbf{water} is found to decrease quite significantly in the upper atmosphere. Decreases can go up to $\sim$68\% after big flare events, and after 10 days of recurring flares changes of up to $\sim$4\% in abundance. These changes occur in the upper atmosphere, i.e. $\geq 85$ km above surface level. Deeper in the atmosphere, the change in water is smaller than 1\% and can be neglected. In contrast, the changes of water in \citet{Tilley2018} are seen almost up until the surface levels. This is because they include an upward flux of water vapor in their models due to the presence of oceans.

\subsection{Emission and Transmission Spectra}

\subsubsection{H$_2$-dominated atmospheres}
\begin{figure}
\hspace{-2.5em}
    \centering
    \includegraphics[scale=0.52]{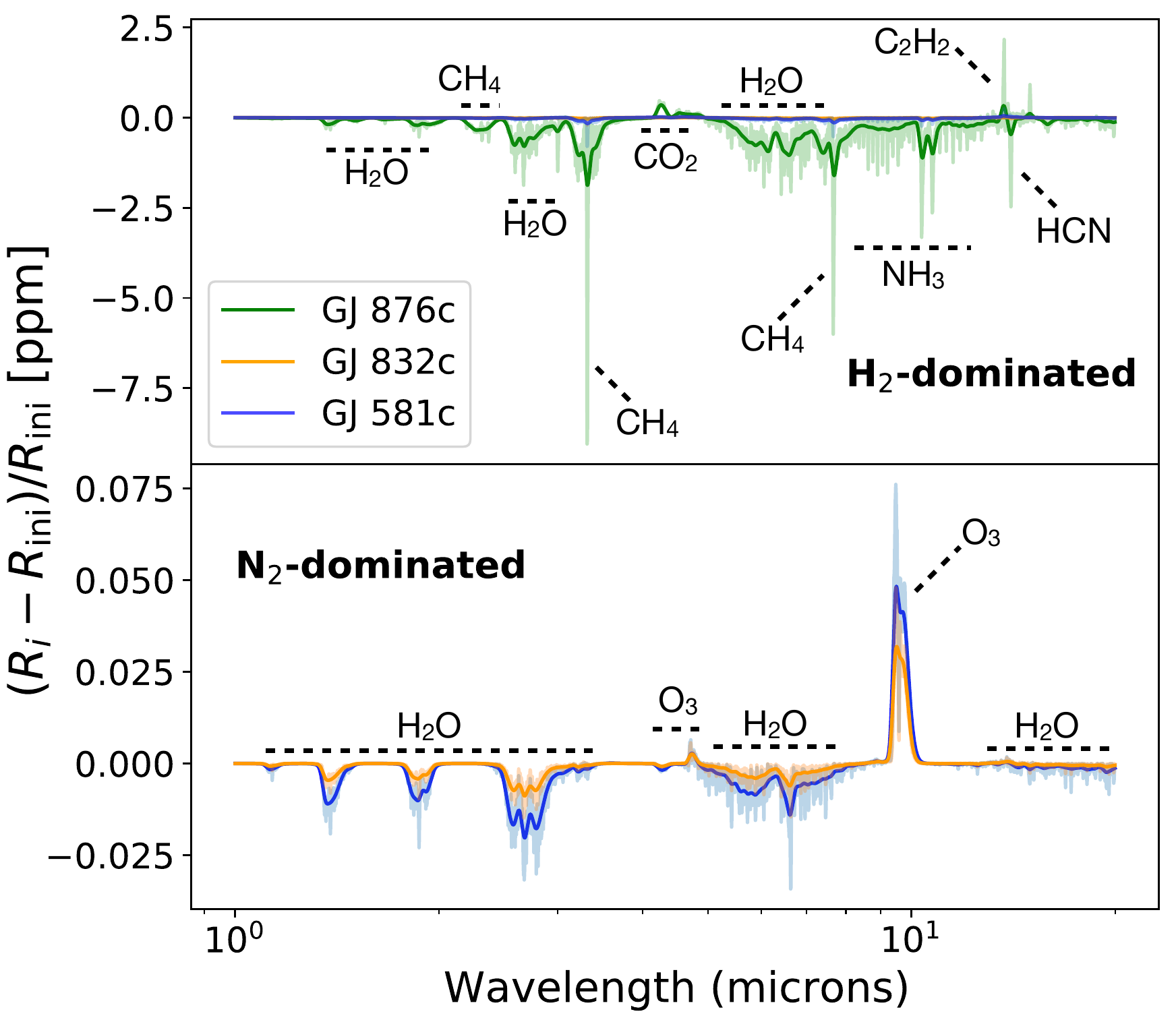}
    \caption{The transmission spectra for all H$_2$-dominated cases (upper panel) and N$_2$-dominated cases (lower panel). The relative change in spectra in ppm is shown in both plots. Here $R_i$ is the radius for a certain wavelength at timestep $i$ (i.e. $t \approx 4.9\times 10^5$ sec), and $R_{\mathrm{ini}}$ is the initial radius where flares are not included in the model yet. The spectra are created at the point where the change in abundance is greatest.}
    \label{fig:emistrans_change_JOV}
\end{figure}

The transmission spectra show a small change over time due to the stellar activity and particle escape, see upper panel in figure \ref{fig:emistrans_change_JOV}. This figure shows the transmission spectra at the point where the abundance change in the molecules are greatest (i.e. after the large flare event) normalized to the initial transmission spectrum. The relative change in transmission spectrum shows largest change for GJ 876c, with a maximum transmission change of $\sim$ 12 ppm at $\lambda \approx 3.314 \mathrm{\ \mu m}$. This feature comes from the photo-dissociation of the molecule CH$_4$. With less methane present, the upper atmosphere becomes less opaque at this particular wavelength, which causes the dip in transmission spectrum.  

\begin{figure*}
    \centering
    \includegraphics[scale=0.73]{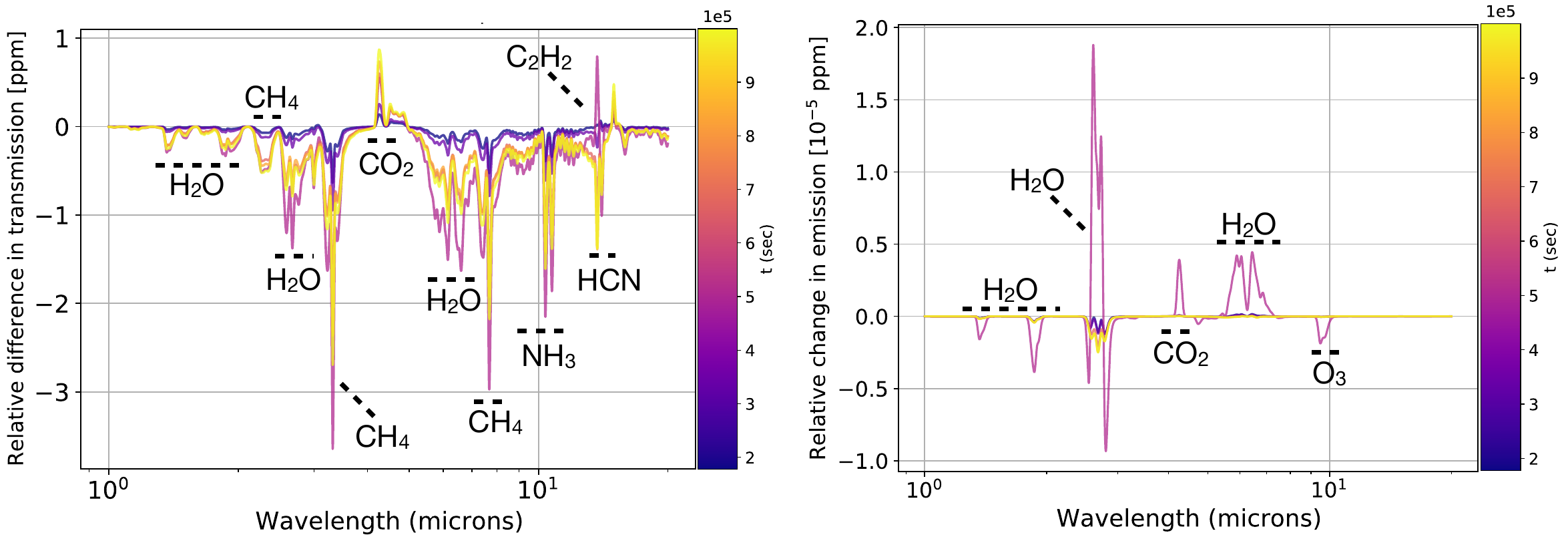}
    \caption{The evolution of the transmission spectra of the most affected H$_2$-dominated atmosphere: GJ 876c (left) and of the emission spectra of the most affected N$_2$-dominated atmosphere in absolute numbers: GJ 581c (right). The color represents the time at which the spectra is created. Relevant features are labelled accordingly. The large flare event occurs around $t \approx 4.9 \times 10^5$ sec and is marked in the colorbar by the white horizontal line. Note that the spectra are convolved with a Gaussian filter to decrease the resolution, which decreases the transmission and emission peaks significantly. }
    \label{fig:trans_GJ581c_evo}
\end{figure*}

The mini-Neptunes, on the other hand, show smaller changes in spectra. The observed radii at $\lambda \approx 3.314 \mathrm{\ \mu m}$ decrease with $\sim 1.1$ ppm and $\sim 0.29$ ppm for GJ 581c and GJ 832c respectively after the big flare event. Where we find that the changes in transmission spectra are non-observable with current technologies, \cite{Venot2016} found that the some features are observable with upcoming space missions such as JWST. Especially the CO/CO$_2$ features showed big changes for the hotter Jupiters ($T_{\mathrm{eff}} = 1303$ K). In this study, on the other hand, we find smaller, insignificant changes for CO$_2$ (at $\lambda = 4.26 \mathrm{\ \mu m}$ in figure \ref{fig:trans_GJ581c_evo}). The main reason for this difference in results is because the planets in \cite{Venot2016} are closer-in to the host star and therefore have higher effective temperatures and suffer more from stellar activity.

Even though the changes in the spectra found here are minimal, over time the spectra seem to change progressively, as can be seen in the left plot of figure \ref{fig:trans_GJ581c_evo}. These figures show the transmission and emission spectra of the seemingly most affected planets in this study. With the exception of the large flare event occurring after $\sim$ 5.7 days ($\sim 4.9 \times 10^5$ sec), the spectra change monotonically over time. Another indicator that the changes in molecular abundances accumulate throughout the simulation. Noticeable is that the spectrum peaks at certain wavelengths at around $\sim 5\times 10^5$ seconds, the moment the largest flare occurs, and recovers moderately afterwards. The dissociation of H$_2$O and formation of C$_2$H$_2$ especially seem very sensitive to these energetic flare events. Nevertheless, it is able to restore a great portion of it. This shows that certain species are able to recover marginally from large flare events, like water (also seen in abundance changes), while other species like methane and ammonia keep on showing these gradual changes. 

Another noticeable feature (from figure \ref{fig:trans_GJ581c_evo}) is the CO$_2$ line at $\lambda = 4.26 \mathrm{\ \mu m}$. Over time, this feature seems to increase linearly. It is expected that this growth remains for a longer period, though at some point it might converge to certain values (e.g. \citeauthor{Venot2016} \citeyear{Venot2016}; \citeauthor{Tilley2018} \citeyear{Tilley2018}). 

\subsubsection{N$_2$-dominated atmospheres}

The lower panel of figure  \ref{fig:emistrans_change_JOV} shows the normalized transmission spectra for the rocky world cases just after the largest flare event. First of all, the changes found in these spectra are, again, quite small as compared with the H$_2$-dominated cases. Nonetheless there are some things worth mentioning about them. At first sight, there are less features to be seen in these normalized spectra as compared to the H$_2$-dominated cases. The transmission spectra, however, do show a very clear feature of O$_3$ at $\sim 9.48 \mathrm{\ \mu m}$, with a maximum of $\sim 0.08$ ppm change in spectrum on GJ 581c. Since the large flare event in the upper atmosphere depletes ozone to a certain level, this peak in transmission must come from deeper levels in the atmosphere where the amount of ozone increases. For both planets this occurs mostly around $P \sim 10^{-6}$ bar.
Even though the change in abundances seem greatest on GJ 832c (see previous section), the transmission spectrum shows greatest change for GJ 581c, most probably caused by its bigger size and larger orbital period.  

Figure \ref{fig:trans_GJ581c_evo} shows the evolution of the emission spectrum of GJ 832c. In the emission spectrum the water feature (at $\lambda = 2.67 \mathrm{\ \mu m}$) is greatest as opposed to the ozone feature in transmission spectrum. Nonetheless, all features in emission spectrum show insignificant changes over time.

Noticeable is the H$_2$O feature at $\lambda = 2.67 \mathrm{\ \mu m}$. For the big flare event around $4.9\times 10^5$ sec, the emission spectrum seems to increase while at other moments in time it shows a decrease. 

Similarly to what was found for the H$_2$-dominated atmospheres, there is a clear pattern in time in how the emission spectrum evolves. The large flare event can again clearly be seen at $\sim 4.9\times 10^5$ sec, after which the atmosphere seems to recover afterwards quite well. At different points in time, the water feature that is seen in this spectrum seems to accumulate throughout time, though with insignificant values. Nonetheless, for a longer time period this accumulation might be observable, assuming it does not converge at later times. 

\section{Discussion}
\label{sec:discussion}
This research showed that including the flare frequency distributions in simulating the effect of flares on exoplanet atmospheres when considering atmospheric escape reveals a gradual change in the chemistry. Our simulations show the evolution of exoplanet atmospheres for 11 days, which is only a first glimpse of the potentially large change in chemistry and evolution of the atmospheres over time. During the simulations, the change in transmission and emission spectra remain very small and non-observable, yet they do accumulate over time for the H$_2$-dominated atmospheres. When simulating the atmospheres for a longer period of time the accumulation might continue even more, to which at some point the atmospheres even show changing features in their spectra during different observations.  Simulating these atmospheres for a longer period is very time intensive, but a necessary step to understand the evolution of exoplanet atmospheres. Therefore, a next step in these kind of studies would be to model such atmospheres over a longer time period to see how they change on e.g. monthly or yearly basis to get a better understanding of the chemical evolution of exoplanet atmospheres. 

\begin{figure*}
    \centering
    \includegraphics[scale=0.75]{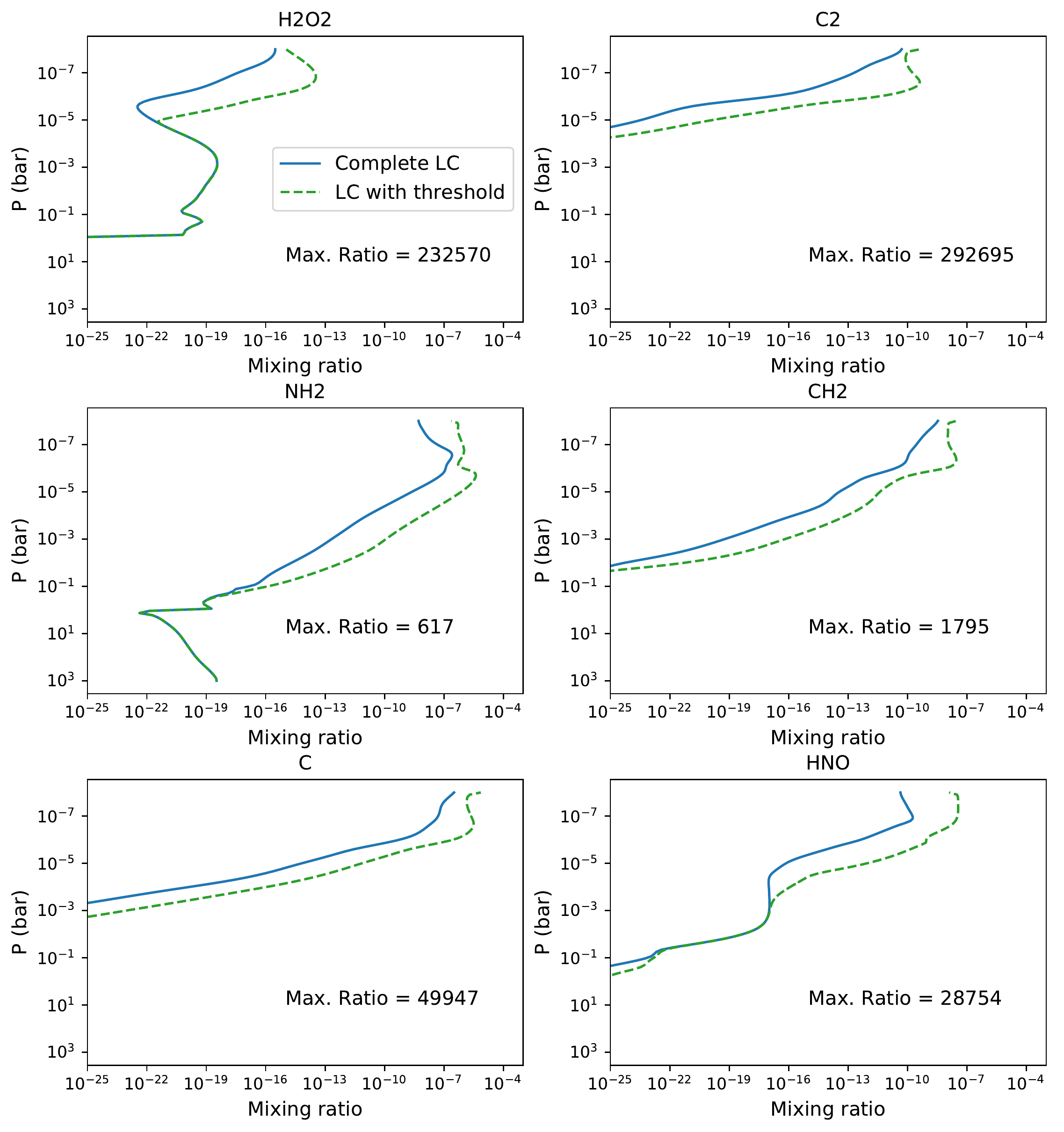}
    \caption{Flare comparison between the two different cases of H$_2$O$_2$, C$_2$, NH$_2$, CH$_2$, C, and HNO, for the many flares case (solid blue) and the less flares case (striped green) on GJ 876c. All curves show the results at the point in time where differences are greatest. The maximum ratios between the two different simulations are given in each plot as the maximum ratio.}
    \label{fig:flare_comp}
\end{figure*}

To reduce the computation times, we also explored the possibility of excluding smaller flares. To test whether setting a threshold in flare energy changes the molecular and atomic abundances significantly over time, we set up two similar runs in VULCAN. In both runs we used the same star- and planet configuration (i.e. GJ 876c), with the exception of the input lightcurve of the stellar spectrum. In one scenario we made use of all flares within the lightcurve while in the other we only included the flares that exceed a certain energy threshold. This threshold is set equal to the mean of all flares within the simulation. Measuring the effect of smaller flares on the final molecular and atomic abundances is done by setting up two similar runs. The final abundance in both cases are then compared with each other, which can be seen in figure \ref{fig:flare_comp} for the most affected species. This figure shows the abundance of both scenarios at the point where the differences are biggest. Overall the results in the two different scenarios show in the upper atmosphere a significant difference in abundance. The smaller flares in general seem to have a great effect in the abundance change whenever a bigger flare occurs afterwards. This might indicate that the effect of the smaller flares accumulate throughout time. This is, however, not the case for all species such as the relevant species like H$_2$, CO$_2$, and H$_2$O, which merely show a maximum ratio of $\sim$1.02 - 3.67.

\begin{figure}
\hspace{-2.5em}
    \centering
    \includegraphics[scale=0.63]{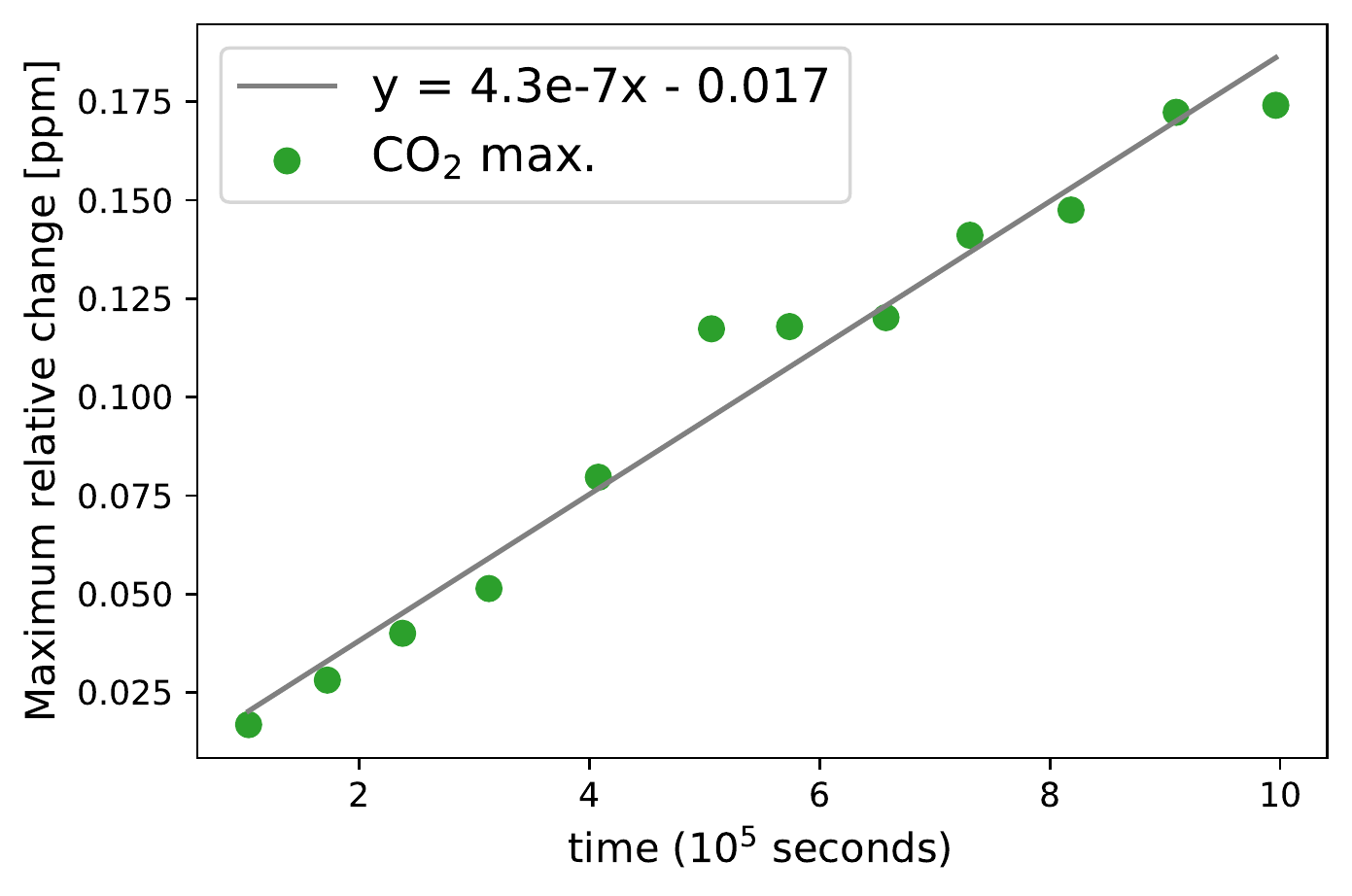}
    \caption{The peak CO$_2$ change (green scatter plot) of transmission spectrum at $\lambda = 4.27 \mathrm{\ \mu m}$ of GJ 876c plotted over time (see CO$_2$ peak in figure \ref{fig:trans_GJ581c_evo}). The grey line represents a linear fit (CO$_2$ peak = 4.3$\times 10^{-7}\cdot$ t (in sec) - 0.017) through the scatter points.}
    \label{fig:spec_extrap}
\end{figure}

The expensive computation times make it challenging to integrate over longer time periods and thus to predict the composition of the variable atmospheres at different points in time. Nevertheless, the accumulating changes in transmission and emission spectra make it possible to estimate the change in spectrum over longer time periods by extrapolating the gradual change. Note, however, that it is expected that the change in spectra converge at some point in time and therefore this extrapolation method can not be used for too big time steps (\citeauthor{Venot2016} \citeyear{Venot2016}; \citeauthor{Tilley2018} \citeyear{Tilley2018}). For the change in CO$_2$ on GJ 876c this is not yet the case (see figure \ref{fig:trans_GJ581c_evo} at $\lambda = 4.27 \mathrm{\ \mu m}$). Figure \ref{fig:spec_extrap} shows the peak value of the change in CO$_2$ as a function of time. In the $\sim$ 11 days of simulating the atmosphere, this peak value keeps on increasing linearly over time due to the recurring flares. When extrapolating this trend it might be possible to find observable changes in spectrum. Using the JWST noise simulator PandExo (\citeauthor{Batalha2017} \citeyear{Batalha2017}), we estimate JWST errorbars using the NIRSpec instrument (G395H mode, R = 2700, f290lp filter). Observing one transit for $\sim$20 hours gives estimated errorbars that extend to approx. 170 ppm in transmission spectrum around $\lambda \approx 4.26 \mathrm{\ \mu m}$. We find that we need to extrapolate the current simulation to $\sim$ 29 years if we want to find spectral changes in CO$_2$ that equal the errorbar sizes of the NIRSpec instrument. There is a possibility that the abundance changes due to recurring flaring have converged within that time-span (\citeauthor{Venot2016} \citeyear{Venot2016}; \citeauthor{Tilley2018} \citeyear{Tilley2018}). Simulating these atmospheres for a bit longer can give more insights on this.

A different caveat to this study is the constant temperature profiles used. The varying spectra from the host stars in the UV- and IR regime could have an impact on the temperature profiles. In this paper we increase the UV-radiation according to the synthetic light-curves and flare spectra. The increase in UV-radiation could subsequently heat up the atmospheres. 
The question remains whether these changes in TP-profiles are significant enough to also alter the chemistry in the atmospheres. 

Due to uncertainties  and lack of observations, we also did not take any lower boundary conditions for the rocky world cases into account. Nevertheless, we point that rocky exoplanets could experience downward and/or upward fluxes of particles due to e.g. volcanoes, sink holes, oceans, and biological processes, which, in turn, influences the composition of the atmospheres (e.g. \citeauthor{Segura2010} \citeyear{Segura2010}; \citeauthor{Tilley2018} \citeyear{Tilley2018}; \citeauthor{Chen2018} \citeyear{Chen2018};  \citeauthor{Yates2020} \citeyear{Yates2020} ). This additional dynamical factor in the chemistry of the atmospheres could impact the lower atmospheres, which, subsequently, has effect on the upper regimes due to vertical mixing. The effect of flaring events could possibly be affected by these chemical changes.

The exclusion of clouds and hazes imposes, in addition, a limitation to this study. The presence of clouds and/or hazes could absorb and scatter UV-light, creating flat observational spectra (e.g. \citeauthor{Brown2001} \citeyear{Brown2001}; \citeauthor{Fortney2005} \citeyear{Fortney2005}; \citeyear{Fortney2013}; \citeauthor{Sing2016} \citeyear{Sing2016}; \citeauthor{Kawashima2017} \citeyear{Kawashima2017}; \citeauthor{Powell2019} \citeyear{Powell2019}; \citeauthor{Ohno2019} \citeyear{Ohno2019}). The effect of photo-chemistry in lower levels of the atmosphere might therefore also be blocked due to their presence. Considering that the temperatures on the planets studied here are low enough for clouds and hazes to form, it would be interesting to see how it impacts the results in future studies. Nevertheless, most relevant changes due to flaring effects found in this study occur in the upper atmosphere, implying that flares impact the atmospheres predominantly above potential cloud decks.  

Finally, we also did not include high energetic particle events emanating from the host star. \cite{Tilley2018} found that the ozone depletion is mostly affected by high energetic particles and not radiation due to flaring. From the results found in this research, N$_2$-dominated planets would be affected mostly by these events which might change the transmission and emission spectra on a greater level. In addition, it is argued in \citet{Chen2021} that coronal mass ejections have a significantly bigger impact on the variable atmosphere than the large flare events.

\section{Conclusion}
\label{sec:conclusion}

 In this study we used synthetic flare spectra to simulate stellar activity of M-dwarf stars and see how this and atmospheric escape impacts the atmospheres of their orbiting exoplanets. The simulation evolved for $\sim$ 11 days, covering a total of 515 flares. We investigated the atmosphere of hydrogen dominated planets as well as nitrogen dominated planets. By implementing a time-dependent stellar activity routine in a chemical kinetics code, we simulated the atomic- and molecular abundances in the atmosphere. Using a radiative transfer code, we also simulated the transmission and emission spectra. Results showed a range of gradual and abrupt changes in abundances and emission and transmission spectra, depending on the species and planetary parameters. To summarize we found for the \textbf{H$_2$-dominated} atmospheres:
 
\begin{itemize}
    \item Overall both gradual and abrupt changes in abundance on all planets due to the recurring flares and atmospheric escape.
    \item Abrupt changes due to flares that can be identified in the abundance evolution plot of OH, H$_2$O, and CH$_4$, with increases/decreases between $\sim 2$-3 orders in magnitude with respect to the initial abundances. The abrupt changes were for some cases (i.e. OH) restored very rapidly, while in other cases (i.e. H$_2$O and CH$_4$) the time of recovery lasted longer than the time interval between subsequent flares. Consequently, changes in these species looked permanent throughout time, with some fluctuations.
    \item Accumulative changes due to flares which can be seen in the species H, H$_2$, OH, and CO$_2$. For H$_2$ this was mostly explained by the atmospheric escape of particles, while for the other species we interpret the results as a combination of both the recurring flares and atmospheric loss of hydrogen. 
    \item An increase in atomic hydrogen, even though the escape rates of atomic hydrogen increased over time. This indicates that the dissociation rate of heavier molecules in to atomic hydrogen is higher than the escape rate.
    \item An accumulative change in transmission spectra  over time, with exception of the largest flare event at $t \approx 4.9\times 10^{5}$ sec. The largest dip in spectrum was seen on GJ 876c for methane at $\lambda = 3.314 \mathrm{\ \mu m}$ with a decrease of $\sim$12 ppm.
   
\end{itemize}

Whereas for the \textbf{N$_2$-dominated} atmospheres we found:

\begin{itemize}
     \item Abrupt changes due to flares that were present for the species N, CH$_4$, H$_2$O, O$_3$, O$_2$, and N$_2$. Nitrogen and ozone showed rapid recoveries to initial abundances, while it took a bit longer for CH$_4$ and H$_2$O to recover. O$_2$ seemed quite resistant to the smaller flare events.
    \item Gradual changes in abundance due to flares found for N, CH$_4$, and O$_3$. Noticeable is that ozone showed an accumulative increase in abundance that went up to an order in magnitude. 
    \item A decreasing trend in the escape rates of both molecular and atomic hydrogen that seemed to converge to lower values over time. This can be explained by the lack of hydrogen in the atmospheres such that the dissociation of heavier molecules in to hydrogen is not enough to sustain high levels in escape rate. 
     \item A gradual change in emission and transmission spectra over time. The maximum increase in transmission spectrum was found on GJ 832c for O$_3$ at $\lambda \approx 4.26 \mathrm{\ \mu m}$ with an increase of $\sim 0.08$ ppm. 
\end{itemize}

In summary, we observed that chemical compositions and emission and transmission spectra change over time due to stellar flares and particle escape. Observing these changes, however, will remain an impossible task on short timescales as evaluated in this study. Nonetheless, this study shows that the atmospheres are dynamical environments. Evolving such atmospheres for longer time periods might even reveal significant changes that could be observed. Current observations of atmospheres of exoplanets could help us understand their history and evolution. 

\section*{Acknowledgements}
We thank the MUSCLES team for in data acquisition of the stellar spectra and their useful insights on the topic. We also thank the referee for the very helpful comments.
\section*{Data availability statement}
The data underlying this article were partly obtained from the MUSCLES survey (https://archive.stsci.edu/prepds/muscles/). All data underlying this article can be obtained upon request to the corresponding author.



\bibliographystyle{mnras}
\bibliography{example} 


\bsp	
\label{lastpage}
\end{document}